## Graphical Abstract

**An efficient GPU-accelerated adaptive mesh refinement framework for high-fidelity compressible reactive flows modeling**

Yuqi Wang, Yadong Zeng, Ralf Deiterding, Jianhan Liang

# Highlights

**An efficient GPU-accelerated adaptive mesh refinement framework for high-fidelity compressible reactive flows modeling**

Yuqi Wang, Yadong Zeng, Ralf Deiterding, Jianhan Liang

- Development of a GPU-accelerated SAMR framework featuring an elaborate time-stepping algorithm with subcycling support and a specialized refluxing algorithm for arbitrary-order Runge-Kutta temporal schemes

- Implementation of a GPU-optimized low-storage explicit Runge-Kutta method for efficient chemical integration, demonstrating significant speedup through reduced register usage

- Comprehensive validation and performance analysis showing remarkable GPU acceleration on both uniform and adaptive meshes, with excellent parallel efficiency across multiple GPU nodes for complex reactive flow simulations

# An efficient GPU-accelerated adaptive mesh refinement framework for high-fidelity compressible reactive flows modeling

Yuqi Wang[a], Yadong Zeng[b], Ralf Deiterding[c], Jianhan Liang[a,*]

[a]*Hypersonic Technology Laboratory, National University of Defense Technology, Changsha, 410073, Hunan, P.R.China*
[b]*Department of Computer Science, University of Texas at Austin, Austin, 78712, Texas, USA*
[c]*AMROC CFD, Brookweg 167, Oldenburg, 26127, Germany*

**Abstract**

This paper presents a heterogeneous adaptive mesh refinement (AMR) framework for efficient simulation of moderately stiff reactive problems. This framework features an elaborate subcycling-in-time algorithm along with a specialized refluxing algorithm, all unified in a highly parallel codebase. We have also developed a low-storage variant of explicit chemical integrators by optimizing the register usage of GPU, achieving respectively 6x and 3x times speedups as compared to the implicit and standard explicit methods with comparable order of accuracy. A suite of benchmarks have confirmed the framework's fidelity for both non-reactive and reactive simulations with/without AMR. By leveraging our parallelization strategy that is developed on AMReX, we have demonstrated remarkable speedups on various problems on a NVIDIA V100 GPU than using a Intel i9 CPU within the same codebase; in problems with complex physics and spatiotemporally distributed stiffness such as the hydrogen detonation propagation, we have achieved an overall 6.49x acceleration ratio. The computation scalability of the framework is also validated through the weak scaling test, demonstrating excellent parallel efficiency across multiple GPU nodes. At last, a practical application of this GPU-accelerated SAMR framework to large-scale direct numerical simulations is demonstrated by successful simulation of the three-dimensional reactive shock-bubble interaction problem; we have saved significant computational costs while maintaining the comparable accuracy, as compared to a prior uniform DNS study performed on CPUs.

## 1. Introduction

Computational simulation of compressible reactive flows, particularly those involving extreme-condition combustion and detonation phenomena, remains one of the most challenging problems in computational fluid dynamics. These flows are characterized by multiple spatial and temporal scales including strong discontinuities, thin chemical reaction zones as well as stiff chemical kinetics, thereby demanding both high numerical accuracy and substantial computational resources. Direct numerical simulations (DNS) of these phenomena with spatially homogeneous grids usually demands a mesh size of $O(10^8)$ even for a moderate-scale problem, rendering it computationally intractable for most practical applications.

To tackle this problem, adaptive mesh refinement (AMR) techniques have been presented as a powerful solution, by dynamically concentrating grid resolution to where it is most needed instead of using uniform meshes everywhere. Some pioneering work of AMR could be found in [1, 2, 3]. Nowadays, it has been further developed into two main branches: the tree-based AMR [4, 5] and the block-structured AMR (SAMR) [6, 7]. Tree-based methods organize and manage the mesh through a cell-based graded tree structure (commonly quad-tree for two dimensions and oct-tree for three dimensions), where each cell can be recursively subdivided into smaller subcells [8]. This hierarchical structure facilitates straightforward synchronization between coarse and fine cells due to their explicit parent-child relationships, offering certain advantages in memory compression through finer-grained control over mesh refinement by avoiding unnecessary refinement and associated memory waste [9]. However, this method also causes inefficiencies in memory accessing due to its cell-based refinement pattern and thus is not suitable for GPU acceleration.

*Correponding author: jhleon@vip.sina.com

In contrast, SAMR employs arbitrarily-sized structured grid blocks to achieve adaptive mesh refinement, allowing flexible local resolution adjustment while maintaining the grid's structured nature. This structured approach ensures contiguous memory access patterns, leading to superior parallel efficiency, which makes it particularly well-suited for implementation on modern high-performance computing architectures. The SAMR technique was first proposed for solving hyperbolic partial difference equations by Berger and Oliger [1] with complex rotated refinement meshes and was subsequently simplified by Berger and Collela [2], where the refined patches are required to be aligned with the coarse mesh. It now has been extensively used for solving PDEs-related problems [10, 11, 12]. A comparison study of tree-based AMR and SAMR approaches in solving partial differential equations (PDEs) could be found in [13].

Nowadays, the advent of Graphics Processing Units (GPUs) has revolutionized scientific computing by offering massive parallelism at relatively low cost. The single instruction multiple threads (SIMT) parallelization manner of GPUs, combined with their high memory bandwidth, provides striking floating-point computation capacity that is particularly well-suited for SAMR computations, in which the calculation is organized in several large blocks of Cartesian uniform meshes. However, efficiently implementing SAMR algorithm for reacting flows on GPUs remains challenging, particularly in retaining the efficiencies of both SAMR algorithm and parallel computing simultaneously. Due to these challenges, previous parallel SAMR solvers [6, 14, 15] and their applications in reactive flow simulation [16, 17] were mostly based on CPUs by using MPI or OpenMP. Although GPU-based solvers for reacting flows have gained increasing attention in recent research [18, 19, 20], studies combining the SAMR algorithm with GPU computing capabilities remain relatively scarce [21, 22]. Therefore, to address these challenges, this work aims to develop a GPU-accelerated parallel SAMR solver for reacting flows simulation based on AMReX [7], an open-source modern high-performance computing framework, which is characterized with flexible block-structured adaptive mesh refinement capabilities and efficient memory management strategies.

In reacting flow solvers, the operator-splitting method is a common implementation to decouple the governing equations into respectively a hydrodynamic component and a chemistry component. The chemistry component, which is essentially integration of an ordinary differential equation (ODE) system, could account for 90% or more of the total simulation time if a very stiff chemical kinetics is chosen [23], implying itself the primary computational bottleneck for large-scale parallel simulation. In an effort to alleviate this issue, various chemical acceleration approaches have been proposed: A class approaches is to use reduction methods to simplify the reaction system, thereby mitigating numerical stiffness [24, 25, 26, 27]. However, for hydrogen-oxygen combustion which is also the focus of this study, the reaction system typically involves only 9 to 13 species and fewer than a hundred of reaction steps. Given the already compact nature of this skeletal mechanism, each intermediate radical and reaction step plays a critical role, rendering the aforementioned model reduction techniques impractical.

Beyond reduction methods, there also exist tabulated and In-Situ-Adaptive tabulated chemistry approaches [28, 29, 30, 31] by pre-computing or dynamically tabulating chemical integration in a high-dimensional space and subsequently retrieve it during calculations. However, the effectiveness of these methods heavily relies on the accuracy of interpolation techniques, and their applicability is limited when dealing with complex multi-physics simulations that involve multiple coexisting combustion regimes. In summary, the simulation of hydrogen-oxygen combustion presents additional challenges for conventional chemical acceleration techniques, necessitating alternative strategies to achieve efficient and accurate computations.

To overcome this challenge, this paper particularly focuses on enabling GPU-accelerated chemical integration. Although implicit solvers, such as DVODE and CVODE [32], are widely applied for chemical integrations due to their ability to handle stiff problems with larger integration step sizes compared to explicit methods. Nevertheless, when implemented on GPUs, where multiple grid blocks can execute concurrently on streaming multiprocessors (SMs) with each thread handling an independent ODE integration, numerous studies have demonstrated that explicit methods outperform implicit methods, particularly for moderately stiff or non-stiff chemical kinetics [33, 34, 35]. For example, Niemeyer and Sung [33] achieved a respective 126× and 25× speedup on GPU than a single- and six-core CPU, by using an explicit fifth-order Runge-Kutta-Cash-Karp method. This method was also adopted in Ghioldi and Piscaglia's study [19] and illustrates a 9.3× speedup on a heterogeneous architecture composed of 128 cores and a NVIDIA V100 GPU with respect to the homogeneous system of only 128 cores. Apart from that, Rao et al. [20] used GPU to accelerate the chemical reactions in supersonic combustion, achieving acceleration ratios ranging from 10 to 80× on different chemical models and problem sizes. Stone et al. [36] used the GPU-enabled fourth-order adaptive Runge-Kutta-Fehlberg ODE solver for chemical integration, and achieved an acceleration ratio of 20.2× than the fifth-order implicit DVODE solver on scalar CPU.

This work presents a state-of-art algorithm for high-fidelity parallel simulation of compressible reactive problems with the combination of SAMR and GPU acceleration. We have made several algorithmic contributions through implementation of an elaborate time-stepping scheme for reacting flows simulations with both subcycling-in-time and non-subcycling-in-time approaches, providing flexibility in balancing algorithm complexity and computational efficiency. We have also presented a specialized refluxing algorithm for FV-AMR solvers that maintain strict conservation properties in Method of Lines (MoL) schemes at arbitrary orders of accuracy, ensuring flux consistency across refinement boundaries.

To further exploit thread parallelism, we propose utilizing a low-storage explicit Runge-Kutta method (LSRK) for chemical integration calculations. Specifically, we implement a GPU-accelerated fourth-order, five-stage, low-storage embedded Runge-Kutta method [37] that requires only three registers per integration step, significantly reducing the register usage compared to conventional explicit methods. Given that the maximum number of concurrent threads on streaming multiprocessors (SMs) is primarily constrained by register consumption, and considering the limited register resources on GPUs, our low-storage approach substantially increases the potential thread-level parallelism for chemical integration. To the best of our knowledge, this optimization strategy for chemical integration on GPUs has not been extensively explored in previous literature.

The remainder of this paper is organized as follows: Section 2 presents the governing equations and numerical methodology of the framework, including our treatment of transport properties and chemical kinetics. Section 3 details the core SAMR implementation within the framework, focusing on our time-stepping and conservation-preserving algorithms. Section 4 discusses the parallelization strategies on CPUs/GPUs and the GPU-accelerated low-storage Runge-Kutta solver for chemical integration, followed by a detailed performance analysis on both CPU and GPU. Section 5 validates the solver through a comprehensive series of canonical cases ranging from fundamental to complex reactive flows. Section 6 illustrates the solver's capability in simulating a large-scale three-dimensional problem on multiple GPUs. Section 7 presents the conclusions and perspectives on future work.

## 2. Numerical Methodology

### 2.1. Governing Equations

Generally, the three-dimensional Navier-Stokes equations with the chemical source term $S_{\text{chem}}$ are employed as the governing equations. Following the conservation law, the basic equations reads

$$\frac{\partial U}{\partial t} + \frac{\partial F}{\partial x} + \frac{\partial G}{\partial y} + \frac{\partial H}{\partial z} = \frac{\partial F^v}{\partial x} + \frac{\partial G^v}{\partial y} + \frac{\partial H^v}{\partial z} + S_{\text{chem}} , \qquad (1)$$

where $U$ is the vector of conservative state variables with $(N_{sp} + 5)$ components and $N_{sp}$ is the number of species

$$U = \left[\rho, \rho u, \rho v, \rho w, \rho E, \rho_1, \rho_2, \ldots, \rho_{N_{sp}}\right]^T , \qquad (2)$$

For vectors of convective fluxes, we have

$$F = \begin{pmatrix} \rho u \\ \rho u^2 + p \\ \rho uv \\ \rho uw \\ (\rho E + p)u \\ \rho_1 u \\ \rho_2 u \\ \vdots \\ \rho_{N_{sp}} u \end{pmatrix}, \quad G = \begin{pmatrix} \rho v \\ \rho uv \\ \rho v^2 + p \\ \rho vw \\ (\rho E + p)v \\ \rho_1 v \\ \rho_2 v \\ \vdots \\ \rho_{N_{sp}} v \end{pmatrix}, \quad H = \begin{pmatrix} \rho w \\ \rho uw \\ \rho vw \\ \rho w^2 + p \\ (\rho E + p)w \\ \rho_1 w \\ \rho_2 w \\ \vdots \\ \rho_{N_{sp}} w \end{pmatrix}, \qquad (3)$$

and for the viscous fluxes, they read

$$F^v = \begin{pmatrix} 0 \\ \tau_{xx} \\ \tau_{xy} \\ \tau_{xz} \\ u\tau_{xx} + v\tau_{xy} + w\tau_{xz} - q_x \\ -J_{x,1} \\ -J_{x,2} \\ \vdots \\ -J_{x,N_{sp}} \end{pmatrix}, \quad G^v = \begin{pmatrix} 0 \\ \tau_{yx} \\ \tau_{yy} \\ \tau_{yz} \\ u\tau_{yx} + v\tau_{yy} + w\tau_{yz} - q_y \\ -J_{y,1} \\ -J_{y,2} \\ \vdots \\ -J_{y,N_{sp}} \end{pmatrix}, \quad H^v = \begin{pmatrix} 0 \\ \tau_{zx} \\ \tau_{zy} \\ \tau_{zz} \\ u\tau_{zx} + v\tau_{zy} + w\tau_{zz} - q_z \\ -J_{z,1} \\ -J_{z,2} \\ \vdots \\ -J_{z,N_{sp}} \end{pmatrix}. \quad (4)$$

where the notation of $\tau$, $J_i$ and $q$ will be given subsequently.

*2.1.1. Perfect gas*

For perfect gas, or calorically perfect gas, the solution vector will degenerate into a simplified version with only five components

$$U = [\rho, \rho u, \rho v, \rho w, \rho E]^T, \quad (5)$$

and thus the last $N_{sp}$ terms in Eq. 3 and 4 also disappear. In this formulation, we do not consider the chemical source term. Since the number of equations about $\rho, u, v, w, E$ is less than the unknowns $\rho, u, v, w, E, p, T$, the solving system is then enclosed by the ideal gas equation and equation of state (EOS).

$$p = \rho RT, \quad (6)$$

$$E = \frac{p}{\gamma - 1} + \frac{1}{2}|\mathbf{v}|^2, \quad (7)$$

where $R$ is the averaged molar weight of the perfect gas, $\gamma$ is the specific heat ratio and $\mathbf{v}$ is the vector of velocities. This formulation is trivial but particularly important for verification of our developed SAMR solver, as used in Section 5.1 and Section 5.3.2 - 5.3.4.

Among the existing viscous fluxes, for the viscous stresses $\tau$, we have

$$\tau = (\mu_B - \frac{2}{3}\mu)(\nabla \cdot \mathbf{v})\mathbf{I} + \mu\left[\nabla\mathbf{v} + (\nabla\mathbf{v})^T\right], \quad (8)$$

where $\mu_B$ and $\mu$ are the bulk and shear viscosity.

For the heat conduction vector $q$, we have

$$\mathbf{q} = -\lambda \nabla T, \quad (9)$$

where $\lambda$ is the thermal conductivity. For perfect gas, we provide with constant transport model $\gamma = const$ and simplified transport with the viscosity $\mu$ determined from the Sutherland formula. Other transport coefficients are determined with the non-dimensionalized numbers $Pr$ and $Le$.

*2.1.2. Real gas*

For high-fidelity combustion simulation, detailed diffusion model and chemistry kinetics should be considered. Therefore, the entire formulations of Eq. 2 - 4 are employed. To close the system, a real gas EOS is considered with

$$E = \sum_{i=1}^{N_{sp}} Y_i h_i - \frac{p}{\rho} + \frac{1}{2}|\mathbf{v}|^2 = \sum_{i=1}^{N_{sp}} \left(Y_i h_{i0}^f + Y_i \int_{T_0}^{T} c_{pi}(s)ds\right) - \frac{p}{\rho} + \frac{1}{2}|\mathbf{v}|^2, \quad (10)$$

where $T_0$ is the reference temperature, $Y_1$, $h_i$, $h_{i0}^f$ and $c_{pi}$ are respectively the mass fraction, total enthalpy, enthalpy of formation and heat capacity at constant pressure of species $i$, obtained using the NASA 7-coefficient polynomial fits [38].

For real gas, the molecular diffusion flux vector $\boldsymbol{J}_i$ is given by

$$\boldsymbol{J}_i = \rho Y_i \mathbf{v}_i = -\rho \tilde{D}_i d_i - \rho Y_i \mathbf{v}^c \tag{11}$$

with

$$d_i = \nabla X_i + (X_i - Y_i)\frac{\nabla P}{P}, \quad \mathbf{v}^c = \sum_{i=1}^{N_p} \tilde{D}_i d_i, \tag{12}$$

where $X_i$ is the $i$-th mole fraction, $\tilde{D}_i$ is the $i$-th diffusion coefficient rescaled by $Y_i/X_i$ and $d_i$ is the $i$-th diffusion driving force. $\mathbf{v}^c$ is the defined correction velocity to ensure mass conservation.

The viscous stress tensor $\tau$ remains still as Eq. 8 and the heat flux vector $\boldsymbol{q}$ is modified by considering the contribution of molecular diffusion

$$\boldsymbol{q} = -\lambda \nabla T + \sum_{i=1}^{N_{sp}} \boldsymbol{J}_i h_i, \tag{13}$$

For determination of detailed transport coefficients ($\lambda$, $\mu$ and $D_i$), the mixture-averaged transport model [39] is employed for its lower computational overhead by neglecting the cross-diffusion effects among different species.

For chemical reactions, the reaction source term $\boldsymbol{S}_{\text{chem}}$ reads

$$\boldsymbol{S}_{\text{chem}} = \left(\dot{\omega}_1, \ldots, \dot{\omega}_{N_{sp}}, 0, 0, 0, 0\right)^T, \tag{14}$$

where the mass production rates $\dot{\omega}_k\left(\rho_1, \ldots, \rho_{N_{sp}}, T\right)$ are derived from a detailed reaction mechanism that consists of $J$ chemical reactions

$$\sum_{k=1}^{N_{sp}} v_{jk}^f S_k \rightleftharpoons \sum_{k=1}^{N_{sp}} v_{jk}^r S_k, \quad j = 1, \ldots, J \tag{15}$$

with $v_{ji}^f$ and $v_{ji}^r$ the stoichiometric coefficients of species $S_k$ appearing as a reactant and as a product. The entire molar production rate of species $S_k$ per unit volume is then given by

$$\dot{\omega}_k = \sum_{j=1}^{J} \left(v_{jk}^r - v_{jk}^f\right)\left[k_j^f \prod_{k=1}^{N_{sp}}\left(\frac{\rho_k}{W_k}\right)^{v_{jk}^f} - k_j^r \prod_{k=1}^{N_{sp}}\left(\frac{\rho_k}{W_k}\right)^{v_{jk}^r}\right], \quad k = 1, \ldots, N_{sp}, \tag{16}$$

with $k_j^f(T)$ and $k_j^r(T)$ denoting the forward and backward reaction rate of each chemical reaction, respectively. Each forward reaction rate is given by an Arrhenius formula and the backward reaction rate is calculated from the respective chemical equilibrium constant.

In this study, we neglect the cross-diffusion terms, i.e., the Soret and Dufour effects, external body forces as well as the the radiant heat transfers for both perfect gas and real gas formulations, due to their minor effects on the simulation results [40, 41].

### 2.2. Numerical Methods
#### 2.2.1. Operator splitting

Due to the stiffness possibly introduced by composing a detailed chemistry model, it is hard to solve the whole inhomogeneous governing system of Eq. 1 for real gases directly. Consequently, we utilize the Strang operator splitting methodology [15] to handle the stiffness of the governing system without compromising the formal accuracy of numerical schemes.

#### 2.2.2. Hydrodynamic part

For the hydrodynamic part, both unsplit finite-volume and dimensional splitting methods are implemented. The complete PDE system is solved with the method of lines (MoL), with the spatial scheme extended to higher order by the Monotonic Upstream-centered Scheme for Conservation laws (MUSCL) reconstruction in each dimension.

$$\begin{aligned}
\tilde{U}_{j+\frac{1}{2}}^L &= U_j^l + \frac{1}{4}\left[(1-\omega)\Phi_{j-\frac{1}{2}}^+ \Delta_{j-\frac{1}{2}} + (1+\omega)\Phi_{j+\frac{1}{2}}^- \Delta_{j+\frac{1}{2}}\right], \\
\tilde{U}_{j-\frac{1}{2}}^R &= U_j^l - \frac{1}{4}\left[(1-\omega)\Phi_{j+\frac{1}{2}}^- \Delta_{j+\frac{1}{2}} + (1+\omega)\Phi_{j-\frac{1}{2}}^+ \Delta_{j-\frac{1}{2}}\right].
\end{aligned} \tag{17}$$

During reconstruction, a slope limiting procedure is made avail to ensure the total variation diminishing (TVD) property with four types of limiters included, i.e., the minmod, superbee, van Leer and van Albada limiters. In Eq. 17, with the weight coefficient $\omega = 0$, a second-order accurate MUSCL scheme is achieved while a spatially third-order accurate MUSCL scheme would be achieved if $\omega = 1/3$. The formal convergence order of the employed schemes is verified in Section 5.1.1.

After reconstruction of the cell interface variables, the Godunov-type Harten-Lax-van Leer-Contact (HLLC) scheme and the FVS-type Advection Upstream Splitting Method (AUSM) scheme are respectively considered to solve the convective fluxes. For viscous fluxes, a commonly-used second-order accurate central difference scheme is employed.

To ensure the overall accuracy of the high-order spatial scheme, the Strong-stability-preserving Runge-Kutta (SSPRK) temporal scheme are utilized. In this study, both SSPRK(2,2) and SSPRK(3,3) [42] are considered.

$$\begin{aligned} U^{(1)} &= U^n + \Delta t f(t^n, U^n), \\ U^{n+1} &= \tfrac{1}{2} U^n + \tfrac{1}{2} U^{(1)} + \tfrac{1}{2} \Delta t f\left(t^n + \Delta t, U^{(1)}\right) \end{aligned} \quad (18)$$

and

$$\begin{aligned} U^{(1)} &= U^n + \Delta t f(t^n, U^n), \\ U^{(2)} &= \tfrac{3}{4} U^n + \tfrac{1}{4} U^{(1)} + \tfrac{1}{4} \Delta t f\left(t^n + \Delta t, U^{(1)}\right), \\ U^{n+1} &= \tfrac{1}{3} U^n + \tfrac{2}{3} U^{(2)} + \tfrac{2}{3} \Delta t f\left(t^n + \tfrac{\Delta t}{2}, U^{(2)}\right). \end{aligned} \quad (19)$$

For the first-order Godunov scheme, it is simply combined with the forward Euler method. For determination of the time step, we use the stability analysis and the corresponding method for multi-dimensional Navier-Stokes equations as given in [14].

*2.2.3. Reactive part*

For the reactive part, integration of the chemical source terms can be essentially regarded as an ODE initial value problem (IVP)

$$\frac{d\hat{U}}{dt} = S_{\text{chem}}, \quad \hat{U} = \left[\rho Y_1, \rho Y_2, \ldots, \rho Y_{N_{sp}}, T\right]^T. \quad (20)$$

Typically, implicit coupled methods have been the preferred choice for solving multi-species systems due to numerical stiff. However, while these methods demonstrate excellent performance on sequential architectures, their efficiency may not translate well to highly parallel systems such as GPUs. Compared to hydro-carbon systems, for hydrogen-oxygen reactions which typically involve less numbers of reaction species ($9 \leq N_{sp} \leq 13$) and reaction steps (fewer than 100), the numerical stiffness is relatively small. This characteristic suggests that explicit methods, when properly parallelized, may actually outperform coupled methods for two main reasons. First, the massive parallelism can effectively compensate for the additional integration steps that are required. Second, implicit coupled methods incur substantial warp divergence from operations such as Jacobian matrix evaluation and non-linear equation system solutions. Indeed, recent studies have demonstrated the superior performances of explicit methods over implicit approaches for moderately-stiff or non-stiff systems by leveraging the SIMT parallelism [33, 36].

In this study, to further exploit the available GPU parallelism, we propose to use a low-storage variant of adaptive explicit Runge-Kutta (LSRK) methods for simulation of hydrogen-oxygen combustion system. This specifically aims to maximum the resided threads on SMs and thus increase the occupancy rates to better hide the latency. While conventional RK methods with comparable order of accuracy ($3 \leq$ order $\leq 5$) typically require 5-9 temporary arrays stored on the register per thread to finish an integration step, a fourth-order LSRK method [37] requires only two register vectors ($R_1, R_2$) for storing intermediate variables, as described in Eq. 21

$$\begin{aligned} \text{(Register 1)} \quad & U^{(j+1)} = X^{(j)} + \left(a_{j+1,j}\right) \Delta t F^{(j)}, \\ \text{(Register 2)} \quad & X^{(j+1)} = U^{(j+1)} + \left(b_j - a_{j+1,j}\right) \Delta t F^{(j)}, \\ \text{(Register 2)} \quad & U^{(j+2)} = X^{(j+1)} + a_{j+2,j+1} \Delta t F^{(j+1)}, \\ \text{(Register 1)} \quad & X^{(j+2)} = U^{(j+2)} + \left(b_{j+1} - a_{j+2,j+1}\right) \Delta t F^{(j+1)}, \end{aligned} \quad (21)$$

where $a_{j,j}$ and $b_j$ are the ordinary Butcher coefficients of the scheme. This significant reduction in register requirements is particularly advantageous in register-limited scenarios, as it potentially allows for a threefold increase in the number of concurrent threads that can reside on the streaming multiprocessors (SMs). Consequently, when each thread independently handles its corresponding ODE problem in an extremely register-limited scenario, this approach can theoretically reduce the overall computation time by a similar factor.

It is worth noting that our implementation of the LSRK method in GPU also includes an embedded scheme to enable time step adaptivity. The main scheme achieves fourth-order accuracy, while incorporating a third-order embedded scheme for error estimation, with the global method at fourth order. The details of the embedded schemes could be found in [37]. An error controller based on weighted root-mean-squared (WRMS) errors is considered here with the new step size adjusted by feedback of the integrated errors

$$
\begin{aligned}
h_{\text{new}} &= h \cdot \min\left(h_{\max}, \max\left(h_{\min}, \beta \cdot \left(\frac{1}{\|e\|_{\text{WRMS}}}\right)^{1/4}\right)\right) \\
\|e\|_{\text{WRMS}} &= \left(\frac{1}{N} \sum_{i=1}^{N} \left(\frac{e_i}{\text{rtol} \cdot |u_i| + \text{atol}}\right)^2\right)^{1/2},
\end{aligned}
\tag{22}
$$

where $i$ denotes the index of component in the state vector, the safety factor $\beta = 0.9$, rtol, atol are respectively the relative, absolute error tolerance, and $\|e\|_{\text{WRMS}}$ is the WRMS error. To prevent excessive variation and possible overshooting, $h_{\max}$ and $h_{\min}$ are set to be $5h$ and $\frac{h}{5}$. In this study, we set rtol = 1e-6 and atol = 1e-10.

In summary, this adaptive solver features a total of five right-hand-side (RHS) evaluations per cell per integration step. This solver is highly storage-efficient, as it requires only one additional register to store the error estimate at each stage. Besides that, since both high-order and low-order schemes share the same RHS calculations, no extra computational overhead is introduced. Due to the embarrassing parallel features of the explicit method, the thread divergence between different cells is expected to be very limited. This is illustrated quantitatively in Section 4.2.2

## 3. Block-structured Adaptive Mesh Refinement

### 3.1. Multi-level grid hierarchy

For the basic SAMR framework, a hierarchical grid structure is constructed by comprising multiple levels indexed by $i = 0, \ldots, l_{\max}$, where $i = 0$ denotes the base level. The computational domain at each level $i$ is characterized by the mesh widths $\Delta x_{n,i}$ ($n = 1, 2, 3$ for three dimensions) and a corresponding time step $\Delta t_i$. For a fine level ($i > 0$), both spatial and temporal resolutions are refined by a refinement factor $r_i$ from its next coarser level $i - 1$.

This property is critical for maintaining numerical stability across the entire grid hierarchy. Within the AMReX framework, the aforementioned hierarchical grid structure is realized through a collection of orthogonal rectangular blocks, with the following basic components serving as the building blocks for data organization:

- *Box*: a block-structured set of grid cells at a given AMR level storing the logical coordinates of the grid

- *BaseFab*: an array-like data container based on the *box* structure storing the corresponding data with *Ncomp* components and *Nghst* ghost cells

- *BoxArray*: the assembly of all *Box*es at a given AMR level

- *FabArray*: a collection of *BaseFab*s associated with a given *BoxArray* and a distribution mapping relationship among multiple processes

Before time stepping, we introduce the boundary condition determination for our SAMR structures, which is implemented through the usage of ghost cells as seen in Fig. 1. For simplification, we illustrate by two ghost cells. For physical boundaries, ghost cells outside the computational domain are populated according to prescribed physical boundary conditions. For grid communication, at the fine-fine boundaries where ghost cells in a grid block $\Omega^{1,0}$ at level-1 overlap with another grid block $\Omega^{1,1}$ at the same level, data of the ghost cells in $\Omega^{1,0}$ is directly populated from the overlapping real cells in $\Omega^{1,1}$. While for coarse-fine boundaries, when ghost cells in a grid block $\Omega^{1,0}$ at

level-1 have no overlapping grid blocks at the same level, they must lie within a particular grid block (here is the $\Omega^0$ at level-0), guaranteed by the proper nesting constraint. In this case, spatiotemporal interpolation is required to populate the ghost cells in the fine grid $\Omega^{1,0}$ from the coarse grid $\Omega^0$, which will be discussed in Section 3.4 in details. Fig. 1 also illustrates the general criterion that within the same level, the real grids may be adjacent but must not overlap, while real grids in adjacent levels may overlap with each other.

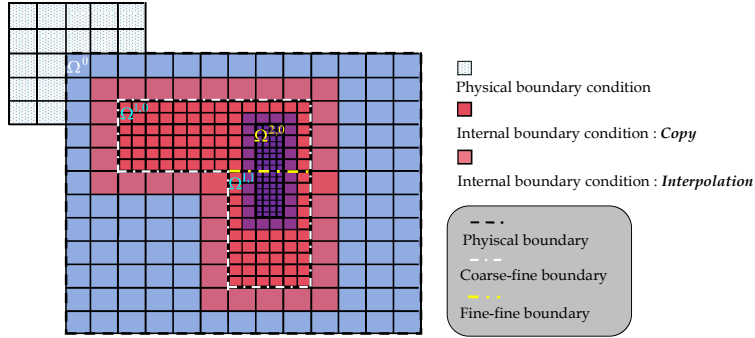

Figure 1: Three-level schematic of the ghost-cell injection on different boundaries in SAMR.

## 3.2. Time stepping strategies

For AMR simulations, we implemented the subcycling-in-time advancement strategy, following the well-established approaches in the literature [1, 2]. This approach has become standard in many production AMR frameworks including Overture [43], Paramesh [44], and AMROC [6] due to its computational efficiency. The subcycling approach advances the solution level by level with hierarchically refined time steps, maintaining a constant time-space ratio throughout the grid hierarchy. This strategy ensures CFL stability while optimizing computational resources by taking appropriately sized time steps at each refinement level. The method minimizes unnecessary computations on coarser grids, which would otherwise be forced to advance at the restrictive time step of the finest level.

The corresponding algorithms are summarized in Algorithm 1.

**Algorithm 1** Subcycling-in-time Advancement
---
1: **function** SUBCYCLING-IN-TIME(*lev*, *time*, *iteration*)
2:     STEPATLEVEL(*lev*, *time*, *dt*[*lev*], *iteration*)               ▷ including refluxing and optional chemical integration
3:     **if** *lev* < *finest_level* **then**
4:         **for** $i \leftarrow 1$ **to** *nsubsteps*[*lev* + 1] **do**
5:             $t_{new} \leftarrow time + (i - 1) \cdot dt[lev + 1]$
6:             SUBCYCLING-IN-TIME(*lev* + 1, $t_{new}$, *i*)
7:         **end for**
8:         AVERAGEDOWNTO(*lev*)
9:     **end if**
10: **end function**

## 3.3. Flux correction

To address flux inconsistencies at coarse-fine grid interfaces, we extend the flux correction algorithm (refluxing) originally proposed by Berger and Colella [2] to be compatible with multi-stage Runge-Kutta methods by scaling the correction at each stage

$$U_{i,j-1}^{l-1(m)} = \hat{U}_{i,j-1}^{l-1(m)} + W_m \delta f_{i,j-1}^{l-1} \frac{\Delta t_l}{\Delta x_{2,l}}, \qquad (23)$$

where $m$ is the current stage of the Runge-Kutta method and $W_m$ is the stage weight. By doing this, we have managed to maintain strictly the global conservation as illustrated in Section 5.3.2.

Similar approaches have been previously implemented in high-order WENO-based AMR frameworks, as demonstrated by Pantano et al. [45] and Ziegler et al [14], and in the Chombo software package [46].

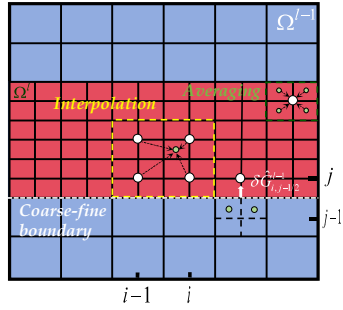

Figure 2: Schematic of refluxing, interpolation and averaging operations.

### 3.4. Coarse-fine synchronization

Data exchange between adjacent levels is managed through conservative averaging and consistent interpolation operations, following standard approaches in AMR methodologies [2, 6, 46].

For averaging, we follow the standard conservative projection

$$\frac{1}{r^D} \sum_{i=1}^{r^D} \phi_f^i = \phi_c. \tag{24}$$

where $r$ is the refinement factor between levels.

For spatial interpolation, we employ linear interpolation (sufficient for second-order FV schemes [2, 6])

$$\phi_f = \phi_c + \sum_{d=1}^{D} \delta_d \cdot \nabla_d \phi_c. \tag{25}$$

For determination of boundary conditions on finer grids, we still need temporal interpolation by using linear interpolation on the coarse time

$$\phi_f\left(t + \kappa \Delta t_f\right) := \left(1 - \frac{\kappa}{r}\right) \phi_c(t) + \frac{\kappa}{r} \phi_c\left(t + \Delta t_c\right), \quad \text{for } \kappa = 0 \ldots r - 1. \tag{26}$$

The implementation of these standard operations offers consistent spatiotemporal second-order accuracy during the recursive hierarchical advancement, as also noted in previous AMR implementations [6, 47].

## 4. Parallelization on CPUs and GPUs

### 4.1. Parallelization strategy

AMReX framework implements a hybrid parallel strategy combining GPU acceleration with MPI-based distributed computing for adaptive mesh computations. Through BoxArray and DistributionMapping abstractions, it enables efficient domain decomposition and dynamic load balancing, while optimizing communication through asynchronous operations and GPU-Aware MPI. The framework's adaptive synchronization mechanisms and optimized GPU computing strategies ensure excellent scalability for large-scale applications.

#### 4.1.1. Hydrodynamic part

In terms of GPU acceleration of hydrodynamic part, it is directly implemented through AMReX's GPU support [7]. Our solver focuses on launching GPU kernels within the AMReX's MFIter (looping all AMReX *Box* at the same level), isolating GPU work to independent data sets on well-established AMReX data objects, ensuring consistency and safety in line with AMReX's coding methodology. To maximize the utilization of GPU parallelism, we further parallelize the applications by utilizing CUDA streams, within which the internal kernels are executed serially while allowing different streams to run simultaneously. This parallelization strategy is illustrated in Fig. 3.

A distributed computing strategy where computationally intensive tasks are offloaded as kernel functions to multiple GPUs through separate MPI processes is employed. Each MPI rank is assigned to a dedicated GPU device, with the inter-process communication is handled at the host side rather than through direct device-to-device communication. Table 1 presents the primary kernel functions during hydrodynamic advancement, which are primarily tailored for dynamic mesh adaptation and time stepping.

Table 1: Primary kernel functions launched on GPU

| Kernel Name | functionality |
| --- | --- |
| *getPhyCellValue* | Converts conserved quantities to primitive variables |
| *calculate_T* | Calculates temperature from conserved quantities using Newton-Raphson iteration |
| *abs_phi* | |
| *fixed_region* | Tag grid cells for refinement based on absolute values, physical location, and gradient criteria |
| *phi_gradient* | |
| *vis_diffcoef_const* | |
| *vis_diffcoef* | Retrieve transport coefficients for viscous terms with constant or variable cofficients |
| *getEdgeValue* | Performs high-order MUSCL reconstruction for cell-interface values |
| *create_numerical_flux* | Calculates numerical hyperbolic fluxes at interfaces using Riemann solver |
| *vis_diff* | Computes viscous fluxes using central difference scheme |
| *TA_unsplit* | |
| *TA_split* | Advances PDEs in time using unsplit or dimensional-splitting methods |

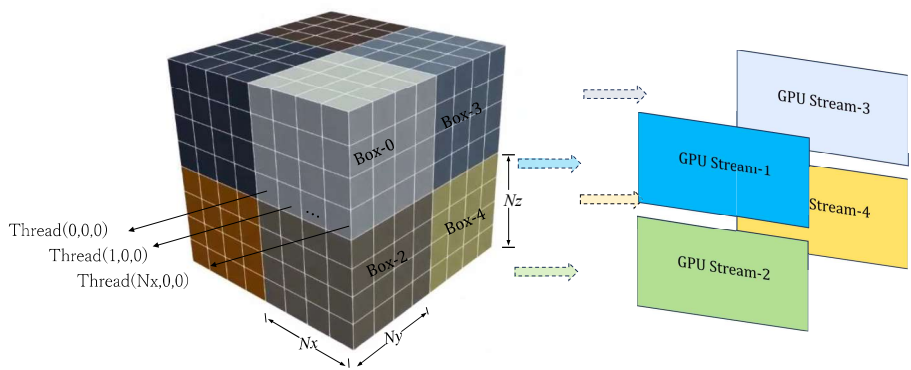

Figure 3: Schematic diagram of the GPU parallelization strategy employed in our solver. In this example, the computational domain is partitioned into $2 \times 2 \times 2 = 8$ *Boxes*, each of which is assigned to a separate CUDA stream to enhance parallelism.

*4.1.2. Reactive part*

In CPU programs, (semi-)implicit methods are commonly used for such stiff ODEs [6, 36], as they offer a larger step size and thus fewer right-hand side (RHS) evaluations. However, this paradigm does not necessarily translate to GPUs, which are designed for high throughput and floating-point computation, but are less efficient at handling complex logical instructions.

Explicit methods are particularly advantageous for GPU computing because they have a minimal logic structure that is friendly to SIMT execution, resulting in fewer conditional branches and reduced risk of warp divergence. Additionally, explicit methods also place less pressure on memory bandwidth, as they do not require storage for Jacobian matrices as in implicit methods. While matrix-free iterative methods such as GMRES [48] can greatly mitigate this issue, implicit methods can be still less efficient due to the necessity of frequent GPU synchronization.

In scope of direct numerical simulations of high-speed combustion [36, 33], the spatial scale employed is usually small enough to capture the turbulence structure, thus making the time step significantly constrained by the stability condition.

Similarly, in the context of AMR with each AMR level advanced serially, the largest portion of the computational elements are concentrated at the highest AMR level. At this level, to resolve small-scale structures, the numerical time step is typically less than $10^{-7}$ s, and for high-speed reactive flows, this value often drops below $10^{-8}$ s. Therefore, we expect the explicit reactive solver to perform similarly good in AMR applications.

Explicit integration of ODEs is inherently embarrassingly parallel, making it a natural fit for GPU acceleration. A parallelization strategy that assigns one thread per ODE has been shown to be more efficient than the one-block-per-ODE approach [36] when the number of ODEs exceeds 1000, achieving up to a 2× speedup as the problem size increases. In practical AMR applications, it typically could provide sufficient parallelism. Nevertheless, a potential limitation of explicit methods in AMR applications is that their relatively coarse-grained parallelism may not fully saturate the GPU when there are possible small boxes generated at the regridding process. To address this issue, we adjust the *max_grid_size* parameter to generate sufficiently large boxes, thereby increasing the number of ODEs solved in parallel. Additionally, by employing multiple CUDA streams, we can further enhance GPU utilization, even in cases where small boxes are created during mesh refinement. These strategies enable us to achieve near-optimal GPU performance with explicit methods, as aligned with our overall parallelization strategy in Section 4.1.1.

It is well known that achieving a high occupancy of streaming multiprocessors (SMs) is essential for improving the GPU parallel efficiency. However, SM occupancy is constrained by several hardware factors, most notably the limited register memory, shared memory, and cache resources. Among these, register availability often becomes the primary bottleneck for memory-intensive kernels. To address this, we employ the LSRK methods. Considering an algorithmic improvement that reduces the number of register vector arrays by three per thread: assume that 1024 threads now reside on each SM, a typical value, though only half of the maximum SM occupancy of the V100; then we could estimate the register savings. For a typical mechanism with 13 species and single-precision arithmetic, reducing three register vectors per thread saves 3*13*4 = 156 bytes per thread. Across 1024 threads, this amounts to 1024*156 = 159,744 bytes of register storage saved. Given that the total register file per SM on the V100 is limited to 256 KB (262,144 bytes), such optimizations are crucial. Without reducing register usage, the kernel may quickly exhaust the available register file, limiting the number of concurrently resident threads and thus reducing occupancy and overall performance. This analysis highlights the necessity of adopting LS RK methods to maximize GPU efficiency for large-scale reactive flow simulations.

## 4.2. Performance analysis
### 4.2.1. Non-reactive case

To evaluate GPU computational performance, we conducted a comparative analysis of kernel computation and communication times for the 2-D Riemann problem described in Section 5.3.4. Tests were performed using both uniform ($N_x \times N_y = 1792 \times 1792$) and structured adaptive mesh refinement (SAMR) grids ($N_x \times N_y = 448 \times 448$ with 3-level refinement, see Section 5.3.4). The comparison was made between pure MPI and pure CUDA implementations on a personal workstation. The pure MPI version is run on an Intel i9-10980XE (3.00GHz) with 18 physical cores and 64 GB memory while the CUDA version is run on a NVIDIA Tesla V100 with 16 GB global memory.

The runtime of each routine is summarized in Table 2 and 3, for the uniform and SAMR grids respectively. For the uniform grid configuration, profiling results revealed remarkable GPU acceleration efficiency, achieving an overall speedup factor of 7.92× compared to the CPU implementation. The hydrodynamic advancement, which constitutes the primary computational workload, achieves a remarkable 7.15× times speedup while showing increased proportional execution time (71.46% to 79.16%) from CPU to GPU implementations. The other kernel with smaller computational intensity also illustrates a 5.12× times speedup, which indicates efficient memory access patterns on GPU.

In contrast, the SAMR case exhibits similar but slightly compromised performance on GPU. The computational efficiency of SAMR is better leveraged in the pure MPI case, as evidenced by the reduced total execution time (19.30s versus 31.16s for uniform grid in pure MPI), while this advantage is less pronounced in the CUDA implementation (3.40s versus 3.93s). It is evident that the overall speedup factor decreases to 5.70×, mostly due to the algorithmic complexity of dynamic mesh adaptation that is allocated to the host only. As can be seen in Table 3, the reduction in parallelization efficiency in GPU in SAMR implementation comes mainly from the coarse-fine communication, whose proportional execution time increases from 9.95% to striking 30.91%. This is attributed to the additional overhead of CPU-only operations in SAMR such as grid partitioning, new mesh creation that cannot be parallelized to GPU. The speedup of hydrodynamic advancement reduces to 5.21×, also due to the additional GPU synchronization overhead that is implicitly contained in each refinement levels.

These results highlight how mesh adaptation strategies impact parallel performance on GPU. While uniform grid computations achieve near-optimal GPU acceleration, the SAMR grid introduces additional computational load due to its complicated algorithmic operations that must be done on the host side. It means for parallel SAMR simulation on GPU, the grid-related operations and frequent synchronizations are the bottleneck that needs to be further optimized.

Table 2: Profiling results for the 2-D Riemann problem on a uniform grid of 1792 × 1792 cells over 10 time steps.

| Routines | Pure MPI Intel i9-10980XE CPU [s] | Pure CUDA 1 NVIDIA Tesla V100 GPU [s] | Speedup |
|---|---|---|---|
| Kernel: hydrodynamic_advancement | 22.27 (71.46%) | 3.112 (79.16%) | 7.15× |
| Communication | 6.7073 (21.54%) | 0.391 (9.95%) | - |
| Kernel: getPhysicalvariable | 2.068 (6.64%) | 0.404 (10.28%) | 5.12× |
| Others | 0.1117 (0.36%) | 0.023 (0.61%) | - |
| Total | 31.1570 (100%) | 3.9300 (100%) | 7.92× |

Table 3: Profiling results for the 2-D Riemann problem on a 3-level SAMR grid with base grid of 448 × 448 cells over 10 time steps.

| Routines | Pure MPI Intel i9-10980XE CPU [s] | Pure CUDA 1 NVIDIA Tesla V100 GPU [s] | Speedup |
|---|---|---|---|
| Kernel: hydrodynamic_advancement | 10.76 (55.75%) | 2.065 (60.67%) | 5.21× |
| Communication | 7.3577 (38.13%) | 1.0512 (30.91%) | - |
| Kernel: getPhysicalvariable | 1.099 (5.70%) | 0.2547 (7.48%) | 4.31× |
| Others | 0.0811 (0.42%) | 0.0321 (0.94%) | - |
| Total | 19.2965 (100%) | 3.4030 (100%) | 5.70× |

*4.2.2. Reaction*

To accurately characterize the spatial variations of reactive flow simulations, we perform this validation by imposing diverse initial conditions on a three-dimensional domain rather than on only one ODE cell. This ODE problem features uniform temperature in the *x*- and *z*-directions and varied temperature distribution in the *y*-direction only, and uniform species composition with $H_2 : O_2 : N_2 = 1 : 2 : 7$. Solution variables include the mass fractions of the chemical species plus the temperature, as indicated by $\hat{U}$ in Eq. 20. To investigate the impact of potential thread divergence on both explicit and implicit solvers, the expression of the temperature distribution in the *y*-direction is designed to be

$$T(y) = T_b + T_r \tanh(k\frac{y - y_m}{L}) + Ae^{-8(\frac{y-y_m}{L})^2} \cos(2\pi f \frac{y}{L}), \qquad (27)$$

where $T_b = \frac{T_{high}+T_{low}}{2}$ represents the averaged temperature, $T_r = \frac{T_{high}-T_{low}}{2}$ indicates roughly the temperature range, $k$ is a steepness factor, $y_m$ is the midpoint position, $L$ is the domain length. $A$ and $f$ are the amplitude and frequency parameters that jointly control the imposed temperature perturbations. Those parameters are further controlled by a smoothness factor $\alpha$, given by

$$k = \frac{6.0}{\alpha}, \quad f = \frac{4.0}{\alpha}, \quad A = \frac{50.0}{\alpha}. \qquad (28)$$

For the smooth profile ($\alpha = 1.0$), the temperature distribution exhibits continuous gradients, while the divergent profile ($\alpha = 0.1$) features large spatial fluctuations. This design deliberately introduces spatially varying numerical stiffness among threads within a warp to induce divergent thread behavior on GPU. In i9-10980, we use the tiling technique to better use the high-speed cache in order to hide the memory access latency by specifying a smaller block size.

To assess accuracy and efficiency, we compare the performances of the low-storage RK54 (LSRK) solver with the implicit CVODE [49] solver and the explicit Runge-Kutta-Fehlberg (RKF) [36] method. In CVODE, we have employed the provided batched sparse QR factorization method by merging the independent ODEs into a large block-diagonal Jacobian matrix with similar sparsity pattern [49]. In the following test results, we have used a single CUDA stream because the box is set large enough to saturate the device. In Fig. 4, the relative error between the LSRK and RKF solutions (the "E2" curve) approaches $10^{-7}$ or lower, due to the similar algorithmic structures. Minor but tolerable differences with CVODE (the "E1" curve) are attributed to the inherent algorithmic differences between explicit and implicit methods. Overall, we have observed roughly 6x and 3x times speedups respectively on the LSRK-GPU solver (0.20 s - 8.93 s) as compared to the CVODE-GPU (1.57 s - 53.15 s) and RKF-GPU solvers (0.55 s - 27.20 s) for a chemical kinetics of 13 species, as seen in Fig. 5. And the effective speedups of LSRK-GPU as compared to its CPU version for both thread-coherence and thread-divergence cases are listed in Table 4 and 5. As can be seen, we have achieved a speedup above 8x times in a size of 65,536 ODEs.

Table 4: Thread-coherence performance of the LSRK integrator, comparison between CPU and GPU.

|            | Problem Size | dt    | Substeps | Wall time [s] | Speedup  |
|------------|--------------|-------|----------|---------------|----------|
| i9-10980XE | 65536        | 1e-05 | 100      | 39.696        | –        |
| V100       | 65536        | 1e-05 | 100      | 4.4032        | 9.0152×  |

Table 5: Thread-divergence performance of the LSRK integrator, comparison between CPU and GPU.

|            | Problem Size | dt    | Substeps | Wall time [s] | Speedup  |
|------------|--------------|-------|----------|---------------|----------|
| i9-10980XE | 65536        | 1e-05 | 100      | 43.237        | –        |
| V100       | 65536        | 1e-05 | 100      | 5.1225        | 8.4406×  |

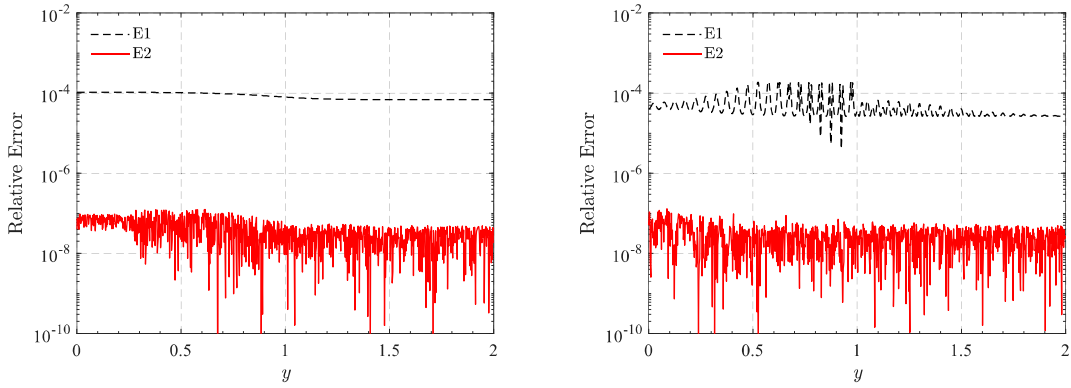

Figure 4: Relative errors of LSRK integrator compared to respectively CVODE (E1) and RKF (E2) solvers in the Left: thread coherence case and Right: thread divergence case.

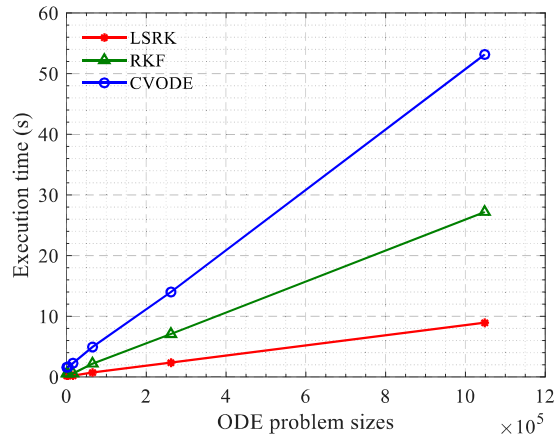

Figure 5: Comparison of execution times among LSRK, RKF and CVODE solvers for the thread-coherence case. Similar results are also observed for the thread-divergence case.

*4.3. Scaling tests*

To evaluate scaling performance of the entire solver, we conduct weak scaling tests on the CNGrid N12 supercomputing system. The GPU partition consists of two different queue configurations. The swarm queue we used consists of 63 computation nodes, with each equipped with Intel Xeon E5-2640 v4 processors featuring 20 physical cores at 2.40GHz and four NVIDIA Tesla V100 GPUs, each with 16 GB HBM2 memory. The nodes in the swarm queue is allocated with 128 GB DDR4 system memory. In our parallel implementation, we assign one MPI process (physical core on the host) per GPU. Therefore, although each node contains 20 physical cores, we only utilize a maximum of four cores (one per GPU) for computation.

The baseline case we employed is the premixed $H_2$-air laminar flame as described in Section 5.2.3, at a grid size of $1024 \times 8 \times 512$ cells (4,194,304 cells in total). For this baseline case, we utilize 1/4 of a node, that is, one MPI rank with one GPU on a GPU partition node. For the weak scaling study, we increase the problem size in the $z$-direction proportionally with the increase of node numbers to maintain a constant computational workload per GPU.

Figure 6 presents the performance from 1 GPU (1/4 node) up to 128 GPUs (32 nodes) on the CNGrid N12 system. The total computational time per step is decomposed into two main components: the flow part and the chemistry part. It is observed that both the flow and chemistry parts of our solver demonstrate nearly ideal parallel efficiency.

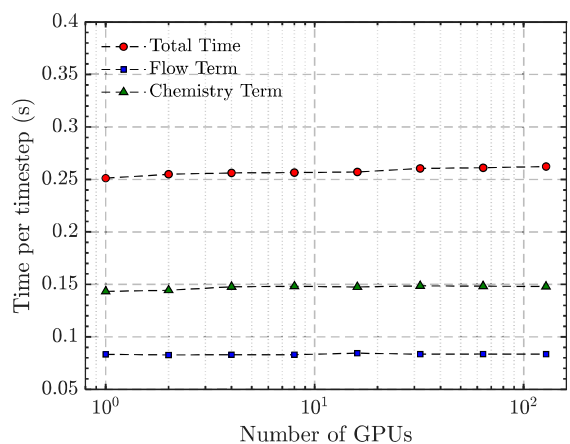

Figure 6: Weak scaling results for the premixed flame problem in Section 5.2.3 with 4,194,304 cells (roughly 80 million Degrees of Freedom) per GPU.

## 5. Benchmark Verification

*5.1. Uniform: perfect gas*

*5.1.1. One-dimensional entropy wave advection*

The entropy wave represents a fundamental test case in computational fluid dynamics, characterized by pure density variations while velocity and pressure fields remain constant. This case serves as an essential benchmark for verifying temporal accuracy, as the wave propagates without deformation and maintains its characteristic shape throughout the simulation. The analytical solution of this case is given by

$$\begin{aligned} \rho &= h(x - ut) + \rho_0 \\ u &= u_0 \\ p &= p_0, \end{aligned} \qquad (29)$$

where $h(x)$ can be any arbitrary function. In this study, we select a sinusoidal perturbation with the following initial condition:

$$[\rho, u, p]_0^T = [1 + 0.2 \sin(2\pi x), 1, 1]^T. \qquad (30)$$

The one-dimensional computational domain spans [-1, 1] in the $x$-direction, equipped with periodic boundary conditions. The exact solution is given by

$$[\rho, u, p]_{ex}^T = [1 + 0.2\sin(2\pi(x - t)), 1, 1]^T. \tag{31}$$

We employ this case to verify the formal order of accuracy for both spatial and temporal discretizations, specifically examining various combinations of Godunov-type schemes and Runge-Kutta time integrators. If not stated, all follow cases run at a CFL number of 0.5. Table 6 presents the convergence analysis using the $L_1$ norm of density error. As demonstrated by the convergence rates, the numerical schemes achieve their designed order of accuracy. The consistent achievement of theoretical convergence rates across different mesh resolutions validates the correct implementation of both spatial and temporal discretization schemes.

Table 6: Grid convergence study demonstrating the order of accuracy for different numerical scheme combinations.

| cells | 1st-order Godunov + forward Euler | | 2nd-order MUSCL + 2nd-order Runge-Kutta | | 3rd-order MUSCL + 3rd-order Runge-Kutta | |
| --- | --- | --- | --- | --- | --- | --- |
| | $L_1$ error of $\rho$ | rate | $L_1$ error of $\rho$ | rate | $L_1$ error of $\rho$ | rate |
| 16  | 0.1698003025   | -      | 0.06736850146   | -      | 0.04283067078   | -      |
| 32  | 0.1330458511   | 0.3519 | 0.01687330103   | 1.9973 | 0.006180214901  | 2.7929 |
| 64  | 0.08507593955  | 0.6450 | 0.004098681478  | 2.0415 | 0.0007914965551 | 2.9650 |
| 128 | 0.04846157189  | 0.8119 | 0.001013603618  | 2.0156 | 9.937298055e-05 | 2.9936 |
| 256 | 0.0259160809   | 0.9030 | 0.0002526104126 | 2.0045 | 1.243260329e-05 | 2.9987 |
| 512 | 0.01340825524  | 0.9507 | 6.310036002e-05 | 2.0011 | 1.55437662e-06  | 2.9997 |

*5.1.2. One-dimensional Sod problem*

Following the verification of smooth wave propagation, we extend our assessment to flows involving discontinuities. The Sod shock tube problem serves as a canonical benchmark for compressible flow solvers, as it simultaneously tests the numerical scheme's capability in resolving shock waves, contact discontinuities, and rarefaction waves.

The one-dimensional computational domain spans a unit length [0, 1], with the initial discontinuity located at $x_d = 0.5$. The initial condition is prescribed as:

$$(\rho, u, p) = \begin{cases} (1, 0, 1) & x \in [0, 0.5] \\ (0.125, 0, 0.1) & x \in (0.5, 1] \end{cases} \tag{32}$$

To evaluate our implementation of the Riemann solver and slope-limiting method, we compare solutions obtained from both the formally first-order and second-order accurate schemes, against a reference solution computed using an exact Riemann solver. Figure 7 presents the density and velocity profiles at the final time of $t = 0.2$ s, with the second-order scheme demonstrating superior performance with reduced numerical dissipation. The excellent agreement with the reference solution validates our implementation of both the approximate Riemann solver and the high-order slope-limiting method. It is noted that we also verify our implementation by successfully reproducing the solution with converse initial setup.

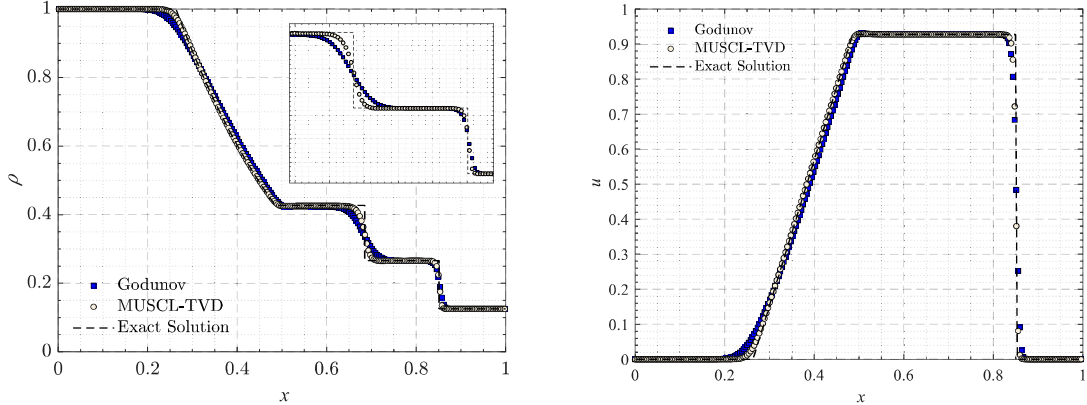

Figure 7: Comparison of numerical solutions using first- and second-order schemes with the exact solution for the Sod shock tube problem.

*5.1.3. One-dimensional viscous shock structure*

This final one-dimensional test case verifies viscous flows. Unlike inviscid shocks with discontinuous properties, viscous shock structures show continuous flow property transitions, helping to verify both convective and diffusive terms. Although this model assumes local thermodynamic equilibrium, it serves as a valuable reference for viscous flow solvers.

The governing equations incorporate the viscous terms of the Navier-Stokes equations:

$$\frac{d}{dx}\begin{bmatrix} \rho u \\ \rho u^2 + p - \tau \\ \rho u H - \tau u + q \end{bmatrix} = 0 \tag{33}$$

where

$$\tau = \frac{4}{3}\mu \frac{du}{dx}, \quad q = -\frac{\mu C_p}{\Pr}\frac{dT}{dx} = -\frac{\gamma \mu}{\Pr(\gamma - 1)}\frac{d(p/\rho)}{dx} \tag{34}$$

The above governing equations are non-dimensionalized through the Rankine-Hugoniot relations. A comprehensive mathematical derivation is presented in [50].

The one-dimensional computational domain spans $[-1.0, 1.0]$ with Dirichlet conditions on both sides. A systematic grid refinement study is conducted with grid spacings of $h = 1/100, 1/200, 1/400$, and $1/800$. The initial condition is constructed from the exact viscous shock solution by integrating Eq. 33 using an explicit four-stage, fourth-order Runge-Kutta method. To comprehensively evaluate our implementation, we examine the constant viscosity model as introduced in Section 2.1.2 with a constant scaled viscosity $\mu^* = 0.14$. Figures 8 present comparisons between numerical and exact solutions for both cases, demonstrating excellent agreement between all flow variables.

Table 7: $L_1$ error norms and convergence rates for the constant viscosity case, demonstrating second-order accuracy across all flow variables.

| h | Density | rate | $x$-velocity | rate | Pressure | rate | Temperature | rate |
|---|---|---|---|---|---|---|---|---|
| 1/100 | 0.0059 | - | 0.0023 | - | 0.0146 | - | 0.0021 | - |
| 1/200 | 0.0020 | 1.5607 | 6.9143e-04 | 1.7340 | 0.0050 | 1.5460 | 7.1342e-04 | 1.5576 |
| 1/400 | 5.3553e-04 | 1.9145 | 1.8173e-04 | 1.9278 | 0.0014 | 1.8365 | 2.0868e-04 | 1.7735 |
| 1/800 | 1.3591e-04 | 1.9648 | 4.6219e-05 | 1.9752 | 3.5106e-04 | 1.9956 | 5.5723e-05 | 1.9049 |

The convergence analysis presented in Table 7 confirms that our spatial discretization of viscous terms achieves the formal second-order accuracy. The convergence rates for all flow variables approach 2.0 as the grid is refined, with the finest mesh ($h = 1/800$) showing rates between 1.90 and 2.0. This systematic convergence behavior is also observed with the application of the Sutherland transport model, thereby fully validating our implementation of the viscous terms in the Navier-Stokes equations.

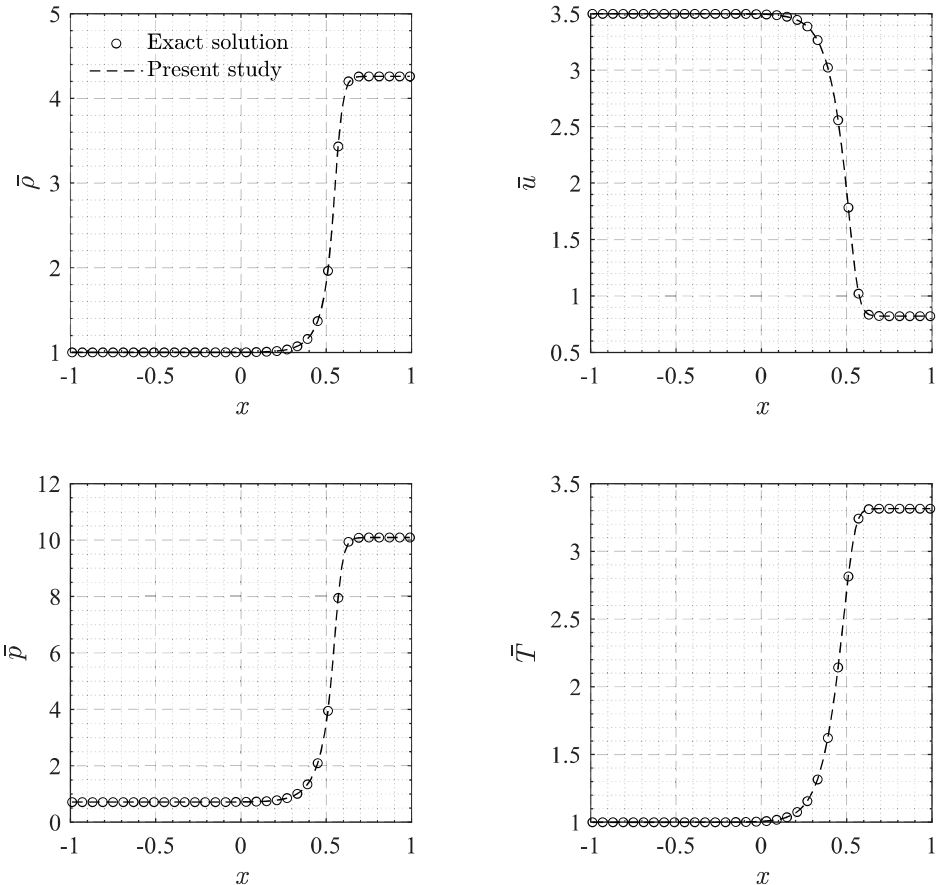

Figure 8: Comparison of scaled density, velocity, pressure and temperature profiles between numerical and exact solutions for the 1D viscous shock example.

#### 5.1.4. Multi-dimensional Method of Manufactured Solutions Verification

Following our verification studies of one-dimensional Euler/Navier-Stokes equations and two-dimensional Euler equations, we present the verification of higher-dimensional cases using the Method of Manufactured Solutions (MMS) [51]. For complex multi-dimensional flow problems where analytical solutions are intractable, MMS provides a rigorous framework for code verification through the construction of analytical solutions with known properties.

The MMS methodology employed in this study follows a systematic approach [52]. First, we prescribe an artificial solution containing sufficient mathematical complexity to exercise all terms in the governing equations. This solution is then substituted into the governing equations to derive corresponding source terms. Finally, these source terms are incorporated into the numerical solver, creating a modified system whose exact solution is known by construction, enabling quantitative assessment of numerical accuracy.

The manufactured solution for our three-dimensional verification is formulated as:

$$\hat{U}(x,y,z,t) = \hat{U}_0 + \hat{U}_x f_s\left(\frac{a_{fx}\pi x}{L}\right) + \hat{U}_y f_s\left(\frac{a_{fy}\pi y}{L}\right) + \hat{U}_z f_s\left(\frac{a_{fz}\pi z}{L}\right) + \hat{U}_t f_s\left(\frac{a_{ft}\pi t}{L}\right), \tag{35}$$

where $f_s$ denotes sinusoidal functions. This choice of trigonometric functions ensures infinite differentiability, facilitates periodic boundary conditions, and generates appropriate coupling between spatial dimensions. The solution parameters are systematically selected: the mean flow state $\hat{U}_0 = [1.2, 1.0, 1.0, 1.0, 20000]^T$ represents $[\rho, u, v, w, p]$, with perturbation amplitudes $\hat{U}_{x,y,z,t} = 0.1\hat{U}_0$ chosen to maintain moderate solution variations. The frequency parameters $a_{fx,fy,fz,ft} = 2.0$ ensure balanced spatial oscillations, while the characteristic length $L = 1.0$ normalizes spatial variations.

Substitution of the manufactured solution of Eq. 35 into the governing equations yields source terms. For example, for the density equation, this term takes the form

$$\begin{aligned}
Q_\rho =& \rho_x \cos\left(\frac{a_{\rho x}\pi x}{L}\right)(u_0 + u_x \sin\left(\frac{a_{ux}\pi x}{L}\right) + u_y \cos\left(\frac{a_{uy}\pi y}{L}\right) + u_z \cos\left(\frac{a_{uz}\pi z}{L}\right))\frac{a_{\rho x}\pi}{L} \\
& - \rho_y \sin\left(\frac{a_{\rho y}\pi y}{L}\right)(v_0 + v_x \cos\left(\frac{a_{vx}\pi x}{L}\right) + v_y \sin\left(\frac{a_{vy}\pi y}{L}\right) + v_z \sin\left(\frac{a_{vz}\pi z}{L}\right))\frac{a_{\rho y}\pi}{L} \\
& + \rho_z \cos\left(\frac{a_{\rho z}\pi z}{L}\right)(w_0 + w_x \sin\left(\frac{a_{wx}\pi x}{L}\right) + w_y \sin\left(\frac{a_{wy}\pi y}{L}\right) + w_z \cos\left(\frac{a_{wz}\pi z}{L}\right))\frac{a_{\rho z}\pi}{L} \\
& + u_x \cos\left(\frac{a_{ux}\pi x}{L}\right)(\rho_0 + \rho_x \sin\left(\frac{a_{\rho x}\pi x}{L}\right) + \rho_y \cos\left(\frac{a_{\rho y}\pi y}{L}\right) + \rho_z \sin\left(\frac{a_{\rho z}\pi z}{L}\right))\frac{a_{ux}\pi}{L} \\
& + v_y \cos\left(\frac{a_{vy}\pi y}{L}\right)(\rho_0 + \rho_x \sin\left(\frac{a_{\rho x}\pi x}{L}\right) + \rho_y \cos\left(\frac{a_{\rho y}\pi y}{L}\right) + \rho_z \sin\left(\frac{a_{\rho z}\pi z}{L}\right))\frac{a_{vy}\pi}{L} \\
& - w_z \sin\left(\frac{a_{wz}\pi z}{L}\right)(\rho_0 + \rho_x \sin\left(\frac{a_{\rho x}\pi x}{L}\right) + \rho_y \cos\left(\frac{a_{\rho y}\pi y}{L}\right) + \rho_z \sin\left(\frac{a_{\rho z}\pi z}{L}\right))\frac{a_{wz}\pi}{L}.
\end{aligned} \quad (36)$$

The analytical source term is then integrated into the numerical solver, as demonstrated by the density field distribution shown in Fig. 9, captured at $t = 0.01$ from the three-dimensional Method of Manufactured Solutions (MMS) verification test case. The computed numerical solution can then be compared with the manufactured exact solution to assess the order of accuracy.

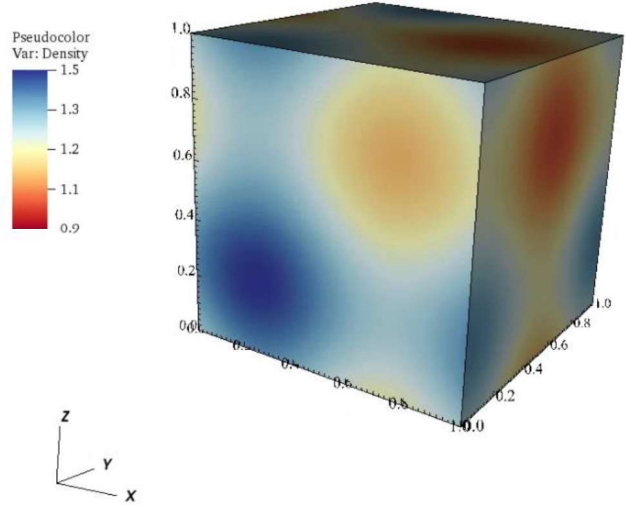

Figure 9: Density Pseudocolor diagram of three-dimensional MMS solution of three-dimensional Navier-Stokes equations.

The numerical scheme employs second-order accurate discretizations in both space and time domains. We consider respectively a unit square and cubic computational domain $[0,1]^2$ and $[0,1]^3$, discretized with uniform grid spacing ($h_x = h_y = h_z = h$) in all directions. Periodic boundary conditions are applied on all domain boundaries to eliminate boundary effects. The simulations are advanced in time with a step size of $\Delta t = 0.1h$, which maintains a CFL number of approximately 0.5, and are carried out until $t = 0.01$.

Both two- and three-dimensional Navier-Stokes equations are verified by MMS, with the verification analysis encompasses multiple quantitative metrics. For the two-dimensional MMS verification, we consider a creeping viscous flow regime characterized by a low Reynolds number (Re = 1) and a very low Mach number ($M = 0.008$). The transport properties are specifically chosen with dynamic viscosity $\mu = 1.0$ and thermal conductivity $k = 2.0$, which ensures solution smoothness while effectively testing the numerical scheme in a diffusion-dominated flow regime.

Conservation properties are rigorously maintained, with mass and energy conservation errors bounded by $10^{-12}$ and $10^{-10}$, respectively. Grid convergence studies demonstrate consistent second-order accuracy, as evidenced by the $L_1$ error norms presented in Table 8. The numerical scheme demonstrates consistent convergence behavior across all variables. The velocity components show approximately second-order convergence with rates of 2.06 ± 0.17 and 2.09 ± 0.14 respectively. The density field exhibits a convergence rate of 1.85 ± 0.11, while the pressure field converges at a rate of 1.56 ± 0.39, showing relatively larger variations in the convergence behavior.

| $h$ | $L_1$ error of $\rho$ | rate | $L_1$ error of $u$ | rate | $L_1$ error of $v$ | rate | $L_1$ error of $p$ | rate |
|---|---|---|---|---|---|---|---|---|
| 1/8 | 0.000245446778960 | - | 0.0285121605000 | - | 0.0281056881488 | - | 1.62292483346 | - |
| 1/16 | 6.7592642529e-05 | 1.8604 | 0.00796281759426 | 1.8402 | 0.00746417404534 | 1.9128 | 0.740545881612 | 1.1319 |
| 1/32 | 2.03704222550e-05 | 1.7304 | 0.0017959481608 | 2.1485 | 0.00163896474398 | 2.1872 | 0.279766743781 | 1.4043 |
| 1/64 | 5.53197259113e-06 | 1.8806 | 0.000402251632482 | 2.1585 | 0.000362752080545 | 2.1757 | 0.0812533995068 | 1.7837 |
| 1/128 | 1.43259184244e-06 | 1.9491 | 9.37783581154e-05 | 2.1007 | 8.43362359202e-05 | 2.1047 | 0.0215773898592 | 1.9129 |

Table 8: Convergence analysis of two-dimensional MMS verification.

While for the three-dimensional MMS verification, we consider a high-Reynolds-number flow regime approaching turbulent conditions, characterized by Re = 100,000. The transport properties are prescribed with relatively small values $\mu$ = 1.0e-05 and $k$ = 2.0e-05, which ensures solution smoothness while effectively testing the numerical scheme in a convection-dominated flow regime. Grid convergence studies demonstrate consistent second-order accuracy, as evidenced by the $L_1$ error norms presented in Table 9. The observed convergence behavior varies across different variables, with density exhibiting an order of 2.20 ± 0.08, velocity components showing orders of 1.57 ± 0.35, 1.50 ± 0.59, and 1.74 ± 0.21 respectively, and pressure demonstrating an order of 2.22 ± 0.09.

| $h$ | $L_1$ error of $\rho$ | order | $L_1$ error of $u$ | order | $L_1$ error of $v$ | order | $L_1$ error of $w$ | order | $L_1$ error of $p$ | order |
|---|---|---|---|---|---|---|---|---|---|---|
| 1/8 | 2.298e-04 | – | 4.385e-03 | – | 4.202e-03 | – | 8.206e-03 | – | 6.263e+00 | – |
| 1/16 | 4.822e-05 | 2.2528 | 1.608e-03 | 1.4475 | 2.525e-03 | 0.7343 | 2.854e-03 | 1.5237 | 1.278e+00 | 2.2926 |
| 1/32 | 1.005e-05 | 2.2623 | 7.066e-04 | 1.1862 | 8.758e-04 | 1.5276 | 9.165e-04 | 1.6388 | 2.607e-01 | 2.2937 |
| 1/64 | 2.228e-06 | 2.1738 | 2.110e-04 | 1.7435 | 2.460e-04 | 1.8317 | 2.526e-04 | 1.8590 | 5.678e-02 | 2.1989 |
| 1/128 | 5.200e-07 | 2.0989 | 5.677e-05 | 1.8941 | 6.461e-05 | 1.9291 | 6.591e-05 | 1.9385 | 1.311e-02 | 2.1153 |

Table 9: Convergence analysis of three-dimensional MMS verification.

These comprehensive verification results provide strong evidence for the correct implementation of our three-dimensional solver. The demonstrated second-order accuracy across all solution components, coupled with robust conservation properties and consistent behavior under varying computational parameters, establishes a solid foundation for subsequent application of the solver to more complex flow problems.

### 5.2. Uniform: real gas
#### 5.2.1. Pseudo One-dimensional multicompoent shock-tube problem

To validate the solver's capability in handling multicomponent shock capturing in three-dimensional domains, a quasi-one-dimensional shock tube problem of dimensions 10 cm × 1 cm × 1 cm is investigated, as shown in Fig. 10(d). The simulation setup consists of an initial discontinuity surface at the midpoint in the $x$-direction, separating two regions with distinct thermodynamic states in Eq. 37. Periodic boundary conditions are imposed at both sides in the $y$- and $z$-directions, while outflow conditions are applied at the $x$-direction boundaries.

$$Q^L = \begin{cases} p_L = 8 \times 10^3 \, \text{Pa} \\ T_L = 400 \, \text{K} \\ u_L = 0 \\ X_{H_2} : X_{O_2} : X_{Ar} = 2 : 1 : 7 \end{cases}, \quad Q^R = \begin{cases} p_R = 8 \times 10^4 \, \text{Pa} \\ T_R = 1200 \, \text{K} \\ u_R = 0 \\ X_{H_2} : X_{O_2} : X_{Ar} = 2 : 1 : 7 \end{cases} \tag{37}$$

The numerical methodology employs a grid of 512 × 2 × 2 cells. A second-order MUSCL-TVD spatial discretization method, together with a second-order Runge-Kutta time integration scheme, is employed. The simulation progresses until reaching a physical time of 4×10$^{-5}$ s, operating at a CFL number of 0.5. The governing Euler equations are solved, utilizing a minmod limiter and HLLC scheme for solving hyperbolic fluxes.

Figure 10(a-c) compare our results with the reference results from Ferrer et al. [40], in which a seventh-order spatial scheme with a third-order temporal scheme was used. The comparison reveals excellent agreement, although our solver achieving lower resolution at the material interfaces due to the lower accuracy of schemes.

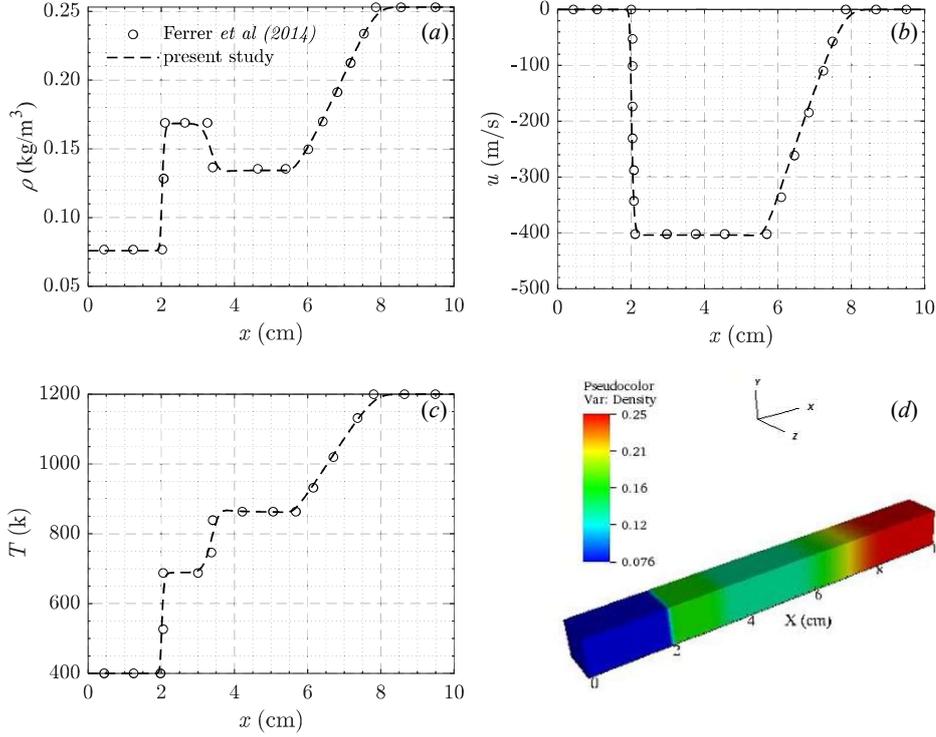

Figure 10: Comparison of flow properties along the centerline between present results (2nd-order MUSCL scheme with RK2 time integration on 512 grid points) and reference data from Ferrer et al. [40] (WENO/CD7 scheme with RK3 time integration on 400 grid points).

*5.2.2. Pseudo One-dimensional Multicomponent Diffusion in Rich $CH_4$/Air Mixture*

Following the validation of inviscid shock-capturing capabilities, we proceed to verify this solver in handling multicomponent diffusion processes. A canonical diffusion problem where a rich methane/air mixture interfaces with a hot exhaust air co-flow is reproduced. This test case, inherently one-dimensional, specifically focuses on validating the implementation of molecular and thermal diffusion terms of real gases in the three-dimensional multicomponent Navier-Stokes equations.

The computational setup maintains a quiescent initial condition with zero velocity and uniform pressure $p = 101325$ Pa throughout the domain. The initial distributions of temperature and species mass fraction are given by

$$\begin{aligned} Y_k(x) &= Y_{k_o} + (Y_{k_f} - Y_{k_o})H(x), \quad k = CH_4, O_2, N_2, H_2O, \\ T(x) &= T_o + (T_f - T_o)H(x), \end{aligned} \quad (38)$$

where the states for both the fuel and oxidizer streams are prescribed according to the configurations described by Vicquelin et al [53]. In consistency with their study, a unity Lewis number assumption is adopted to calculate the molecular diffusion coefficient $D_k$, while $\mu$ and $\lambda$ are determined from the detailed transport model as described in Section 2.1.2.

The numerical domain spans 5 cm × 1 cm × 1 cm, where $x$ is the primary diffusion direction. The computational mesh comprises 256 × 2 × 2 cells, with periodic boundary conditions applied on all boundaries. The numerical scheme we employed is also of second-order accuracy.

Figure 7 presents the computed profiles of temperature and methane mass fraction at $t = 0.5$ s. The numerical results demonstrate excellent agreement with the reference data from Vicquelin et al. [53] in which an eighth-order accurate central difference scheme is used. This agreement validates the correct implementation of the diffusive flux terms in our multicomponent Navier-Stokes solver.

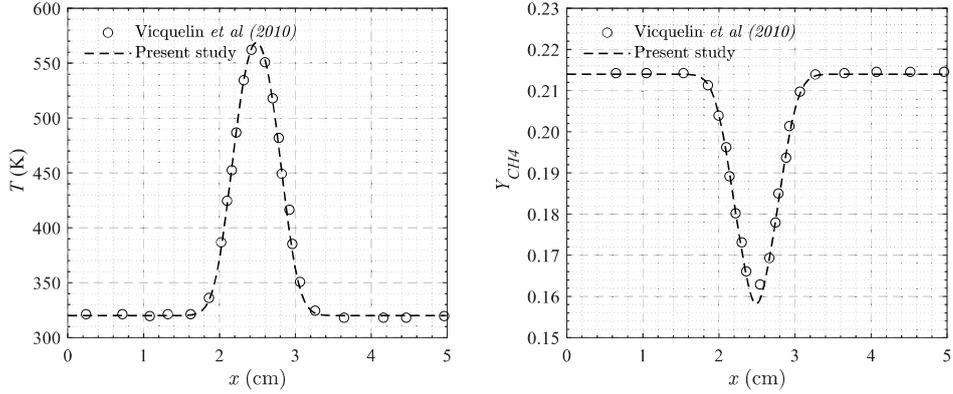

Figure 11: Temperature and $CH_4$ mass fraction profiles for the rich methane/air diffusion case at $t = 0.5$ s, compared with reference data from Vicquelin et al. [53].

### 5.2.3. Pseudo One-dimensional Premixed $H_2$-Air Laminar Flame

Having validated the multicomponent advection and diffusion processes, we proceed to verify the implementation of chemical reaction terms through a canonical reactive case. This verification is crucial as it examines the coupled interaction between convection, diffusion, and chemical reactions in a premixed flame configuration.

The test case simulates a one-dimensional stoichiometric hydrogen-air premixed laminar flame in a three-dimensional computational domain of 1 cm × 1 cm × 1 cm. The chemical kinetics is determined from the detailed mechanism of Burke et al. [54], which consists of 13 species and 27 reversible reactions. The initial flame profile is obtained from a one-dimensional calculation using Cantera [55] under the constant volume (CV) condition, and the laminar flame speed is calculated as 2.33 m/s.

The computational domain is discretized into 512 × 2 × 2 grid points. The initial conditions are set to atmospheric pressure $p = 101325$ Pa with an unburned gas temperature $T = 300$ K at the molar fraction ratio $X_{H_2} : X_{O_2} : X_{N_2} = 2 : 1 : 3.76$. The left boundary is prescribed as an inlet of the unburned mixture with an inflow velocity equal to the laminar flame speed, while the right boundary implements the outflow condition with first-order extrapolation. Periodic conditions are applied in the $y$- and $z$-directions.

To facilitate quantitative comparison with the reference solution, the physical coordinate is transformed into a normalized progress variable $Z$ defined as

$$Z = \frac{T - T_u}{T_b - T_u} \tag{39}$$

where $T_u$ and $T_b$ represent the unburned and burned gas temperatures, respectively. This transformation accounts for the slow drift of the flame position towards the inlet due to pressure waves generated by the heat release.

The numerical methodology combines second-order spatial discretization and the LSRK chemical integrator described in Section ?? through a second-order efficient Strang splitting method given in Eq. ??. The simulation proceeds at a CFL number of 0.4 until it reaches a physical time of $1\times10^{-3}$ s, allowing sufficient time for the flame structure to attain a quasi-steady state.

Figure 12 presents a comparison of the major mass fractions of species and the temperature profiles between our numerical results and the reference solution of Cantera. Both subcycling-in-time and non-subcycling-in-time methods demonstrate excellent agreement with the reference solution, particularly in capturing the steep gradients in the flame front region. This agreement validates the correct implementation of the chemical source terms and their coupling with transport processes in our solver.

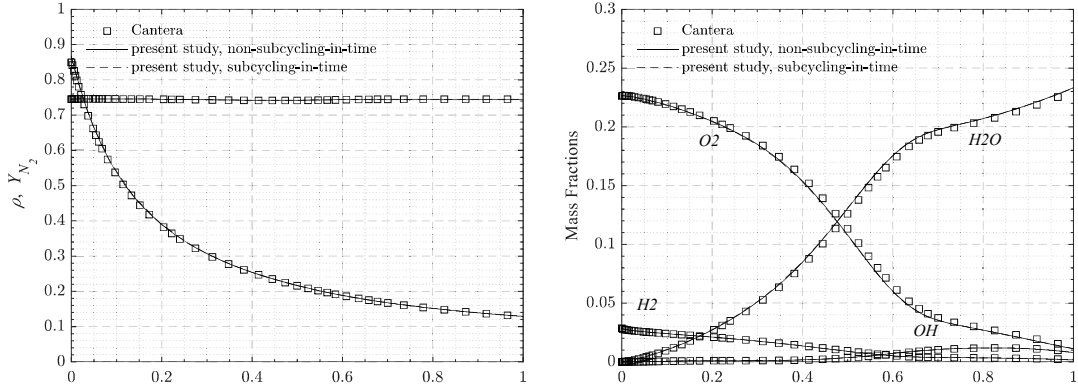

Figure 12: Comparison of temperature and major species mass fraction profiles in premixed stoichiometric $H_2$-air laminar flame: present results with subcycling (solid lines) and without subcycling (dashed lines) versus Cantera reference solution (symbols). The profiles are plotted against the normalized progress variable $Z$.

### 5.3. SAMR

#### 5.3.1. SAMR error estimation

The error estimation methodology for SAMR simulation can be formulated differently from the single-level error estimation. For a single refinement level, the analytical expression of $L_1$ norm calculation in three dimensions is given by

$$L_1(\Delta x, \Delta y, \Delta z, \Omega) = \sum_{i,j,k} |q_{i,j,k} - q_{i,j,k}| \Delta x \Delta y \Delta z, \tag{40}$$

while in multi-level SAMR implementations, the total SAMR error estimation can be expressed as

$$L_1^{\text{SAMR}} = L_1^{\Lambda}(\Delta x_{\Lambda}, \Delta y_{\Lambda}, \Delta z_{\Lambda}, \Omega_{\Lambda}) + \sum_{l=0}^{\Lambda-1} L_1^l(\Delta x_l, \Delta y_l, \Delta z_l, \Omega_l \setminus \Omega_{l+1}), \tag{41}$$

where $\Lambda$ is the current maximum refinement level and $L_l$ ($l < \Lambda$) is computed over the region of level $l$ exclusive of its subsequent finer level $l + 1$.

#### 5.3.2. One-dimensional SAMR entropy wave advection

The structured adaptive mesh refinement (SAMR) framework offers significant advantages in computational efficiency by concentrating computational resources in regions requiring higher resolution. This section revisits the one-dimensional entropy wave advection problem studied in Section 5.1.1 to validate the SAMR implementation. Our investigation of this section focuses on three critical aspects: maintenance of the scheme's formal order of accuracy and effectiveness of the refluxing operation. Comparative performance of different time-stepping paradigms (subcycling and non-subcycling in time) has been shown earlier in Section ??.

The simulation parameters remain consistent with the single-level case, with a simulation time of 1.0 s and a CFL number of 0.5. We examine a two-level refinement configuration to comprehensively evaluate the solver's performance under different mesh hierarchies.

To systematically validate our SAMR implementation, we first verify that the solver still maintenance of the scheme's designed convergence properties from the single-level framework, ensuring that the SAMR infrastructure does not compromise the underlying numerical scheme's accuracy. Table 6 has presented the baseline verification results, confirming that the first- to third-order schemes achieve their theoretical order of accuracy on uniform grids. Following this foundational verification, we conduct a comprehensive analysis using two different refinement strategies, each designed to evaluate specific aspects of the SAMR implementation:

*statically local refinement.* To evaluate the solver's performance with fixed mesh interfaces, we implement static local refinement by restricting the refined region to $[x, y] \in [-0.5, 0.5] \times [-1, 1]$ as shown in Fig. 13. This configuration, with regridding disabled, allows us to isolate and assess the accuracy of interface treatments and flux calculations across refinement boundaries. Table 10 shows that both subcycling and non-subcycling time stepping methods approach the second convergence order (the formal order of accuracy), and the non-subcycling method demonstrates slightly superior convergence rates due to smaller time step taken.

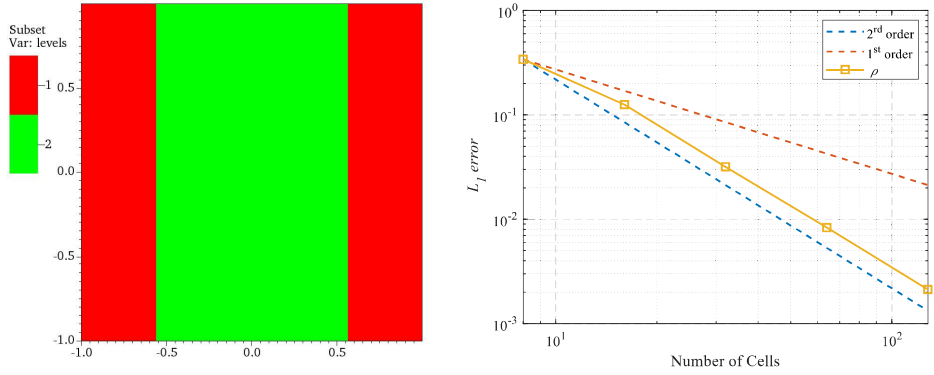

Figure 13: Grid refinement illustration and convergence analysis of the one-dimensional 2-level SAMR entropy wave advection with local static refinement.

Table 10: Convergence study for different stepping methods with static refinement

|     | non-subcycling | | subcycling | |
| --- | --- | --- | --- | --- |
| $N$ | SAMR $L_1$ error | rate | SAMR $L_1$ error | rate |
| 8   | 0.3401625286 | -     | 0.3494165553 | -     |
| 16  | 0.125190705  | 1.442 | 0.1319160502 | 1.405 |
| 32  | 0.0318242950 | 1.976 | 0.03426894159 | 1.945 |
| 64  | 0.0083011802 | 1.939 | 0.00887645923 | 1.949 |
| 128 | 0.0021273074 | 1.964 | 0.002260410424 | 1.973 |

*dynamically local refinement.* The more challenging and realistic SAMR configuration is illustrated in Fig. 14, in which we employ dynamic local refinement approach triggered in this case by the criterion $[x, y] \in [-0.5 + t, -0.2 + t] \times [-1, 1]$ and with a regridding interval of two. This setup evaluates the solver's ability to maintain accuracy and conservation properties under frequent mesh adaptation. As demonstrated in Table 11, our convergence analysis confirms the effectiveness of the interpolation approach, although a slight degradation in convergence rate is observed as the mesh resolution increases. This effect can be effectively mitigated by employing smaller time steps, specifically with a CFL number of 0.1.

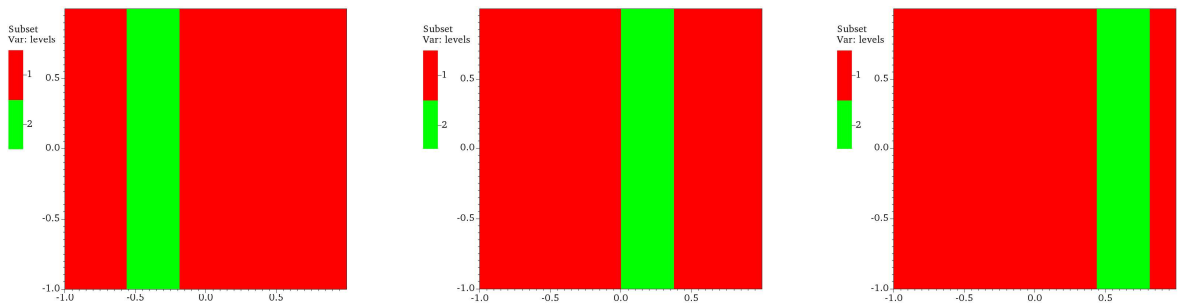

Figure 14: Grid refinement illustration and convergence analysis of the one-dimensional 2-level SAMR entropy wave advection with local dynamic refinement.

Table 11: Convergence study for different stepping methods with dynamic refinement

|  | non-subcycling | | subcycling | |
|---|---|---|---|---|
| N | SAMR $L_1$ error | rate | SAMR $L_1$ error | rate |
| 8 | 0.2948286063 | - | 0.3275755547 | - |
| 16 | 0.1080560931 | 1.448 | 0.1240438937 | 1.400 |
| 32 | 0.0312380594 | 1.790 | 0.03498875927 | 1.825 |
| 64 | 0.009748048552 | 1.680 | 0.008301845899 | 2.075 |
| 128 | 0.003001812537 | 1.699 | 0.002377939946 | 1.803 |
| 256 | 0.0009554229042 | 1.651 | 0.0007293184009 | 1.705 |

Lastly, we evaluate the conservation properties of the conserved quantities $\rho, \rho u$ and $\rho E$ in both statically and dynamically refined meshes, with and without refluxing operation across subcycling and non-subcycling implementations. As observed in Fig. 15, for both subcycling and non-subcycling paradigms, conservation is strictly maintained for all variables in both dynamic and static refinements when refluxing is properly implemented. This validates our implementation of our MoL refluxing approach as presented in Section ??. In contrast, simulations without refluxing exhibit significant conservation errors – manifesting as periodic fluctuations in static refinement cases and, more severely, as monotonically increasing deviations in dynamic refinement scenarios. These conservation violations can lead to substantial numerical artifacts, particularly in applications involving shock capturing and reaction prediction.

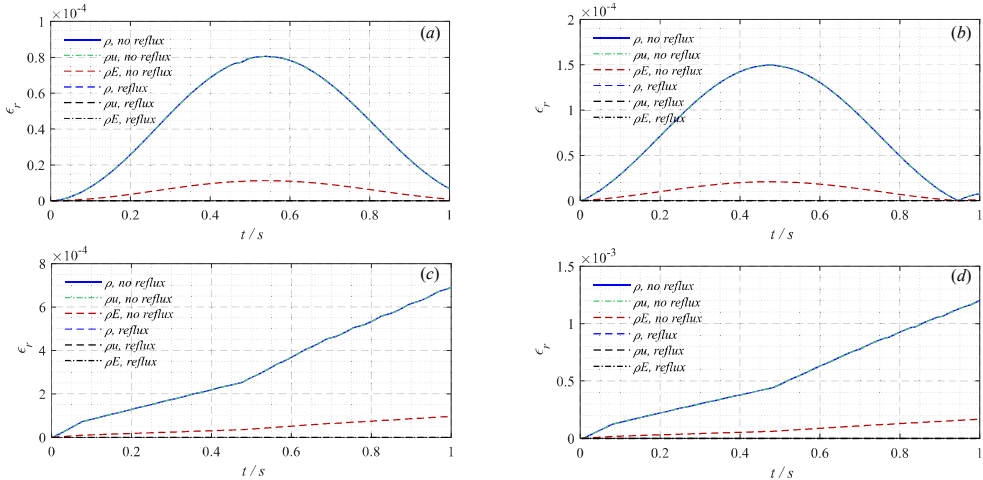

Figure 15: Numerical conservation of the state vector with and without refluxing operation in both subcycling- and non-subcycling-in-time paradigms. (a-b) is non-subcycling and subcycling-in-time with statically refinement, (c-d) is non-subcycling and subcycling-in-time with dynamically refinement.

### 5.3.3. Two-dimensional unsteady isentropic vortex convection

This case is used to verify the accuracy of our solver in solving two-dimensional Euler equations with minor discontinuity. The initial vortex is located at $(x_0, y_0)$ = (-10, -10), convecting with free stream velocity, passing through the origin at $t = 5$, and finally reaching $(x, y) = (10, 10)$ at $t = 10$.

The exact solution reads

$$\frac{u(x,y,t)}{a_\infty} = \frac{u_\infty}{a_\infty} - \frac{\beta}{2\pi a_\infty}\bar{y}e^{\alpha(1-r^2)/2},$$

$$\frac{v(x,y,t)}{a_\infty} = \frac{v_\infty}{a_\infty} + \frac{\beta}{2\pi a_\infty}\bar{x}e^{\alpha(1-r^2)/2},$$

$$\frac{T(x,y,t)}{T_\infty} = 1 - \frac{\beta^2(\gamma-1)}{8\pi^2 a_\infty^2}e^{\alpha(1-r^2)},$$ (42)

$$\frac{\rho(x,y,t)}{\rho_\infty} = \left(\frac{T(x,y,t)}{T_\infty}\right)^{\frac{1}{\gamma-1}},$$

$$\frac{p(x,y,t)}{p_\infty} = \left(\frac{T(x,y,t)}{T_\infty}\right)^{\frac{\gamma}{\gamma-1}}$$

where

$$\bar{x} = x - x_0 - u_\infty t$$
$$\bar{y} = y - y_0 - v_\infty t$$
$$r = \sqrt{\bar{x}^2 + \bar{y}^2}$$

The computational parameters are set as follows: free stream sonic speed $a_\infty = 1$, specific heat ratio $\gamma = 1.4$, free stream velocity $(u_\infty, v_\infty) = (2, 2)$, and vortex characteristic parameter $\alpha = 1, \beta = 5$. The initial solution is set with $\rho_\infty = 1$, $T_\infty = 1$ and $p_\infty = 1$.

The numerical simulation employs a two-level SAMR configuration, with periodic boundary conditions applied in all directions. The mesh hierarchy consists of one refinement level above the base grid, with a refinement ratio of 2. The regridding operation is performed every 2 time steps, guided by the absolute value refinement criteria of density. The numerical scheme employs an unsplit method with second-order spatial accuracy and second-order Runge-Kutta time integration, maintaining a CFL number of 0.5 using a non-subcycling-in-time stepping framework. In the convergence analysis, honogeneous mesh sizes of $h = 1/16, 1/32, 1/64, 1/128$ and $1/256$ are considered.

Figure 10 illustrates the vortex convection through the mainstream flow through the density isolines and grid refinement. As can be seen, the isentropic vortex is properly contained and resolved during its full advection in a maximal two-level refinement. Here, we use a absolute refinement criteria by letting the cells with the density lower than 0.99 tagged for refinement. The refinement diagrams illustrates this criteria works well in this case. Furthermore, the convergence rates of density, x-direction velocity, y-direction velocity and pressure are listed in Table 10. The $L_1$ error is estimated in a SAMR approach. As can be seen, beginning at $h = 1/256$, all variables occur a slight decrease of covergence rate, because the vortex center more and more appears to be a sharp discontinuity with global mesh refinement. This makes the solution no longer smooth anymore, so the error estimation method based on the Taylor series do not strictly apply. However, this situation could be significantly improved with the usage of slope limiter with minimal sacriface on the order accuracy. Notely, since we also enable the refluxing operation the variable vector is conserved through the entire computation.

| $h$ | $L_1$ error of $\rho$ | rate | $L_1$ error of $u$ | rate | $L_1$ error of $v$ | rate | $L_1$ error of $p$ | rate |
|---|---|---|---|---|---|---|---|---|
| 1/16 | 2.833158243 | - | 10.64202206 | - | 10.54875514 | - | 3.98503671 | - |
| 1/32 | 1.854273008 | 0.6116 | 4.678573408 | 1.1856 | 4.737923462 | 1.1547 | 2.251107326 | 0.8240 |
| 1/64 | 0.7272644164 | 1.3503 | 1.850440125 | 1.3382 | 1.887513211 | 1.3278 | 0.9419262961 | 1.2569 |
| 1/128 | 0.185438184 | 1.9715 | 0.4660257159 | 1.9894 | 0.4889227924 | 1.9488 | 0.2436031395 | 1.9511 |
| 1/256 | 0.04192341116 | 2.1451 | 0.1097151939 | 2.0866 | 0.1168223923 | 2.0653 | 0.05583009545 | 2.1254 |
| 1/512 | 0.009898493811 | 2.0825 | 0.02689294632 | 2.0285 | 0.02847226699 | 2.0367 | 0.01338322855 | 2.0606 |

Table 12: Convergence analysis of two-dimensional unsteady isentropic vortex convection problem.

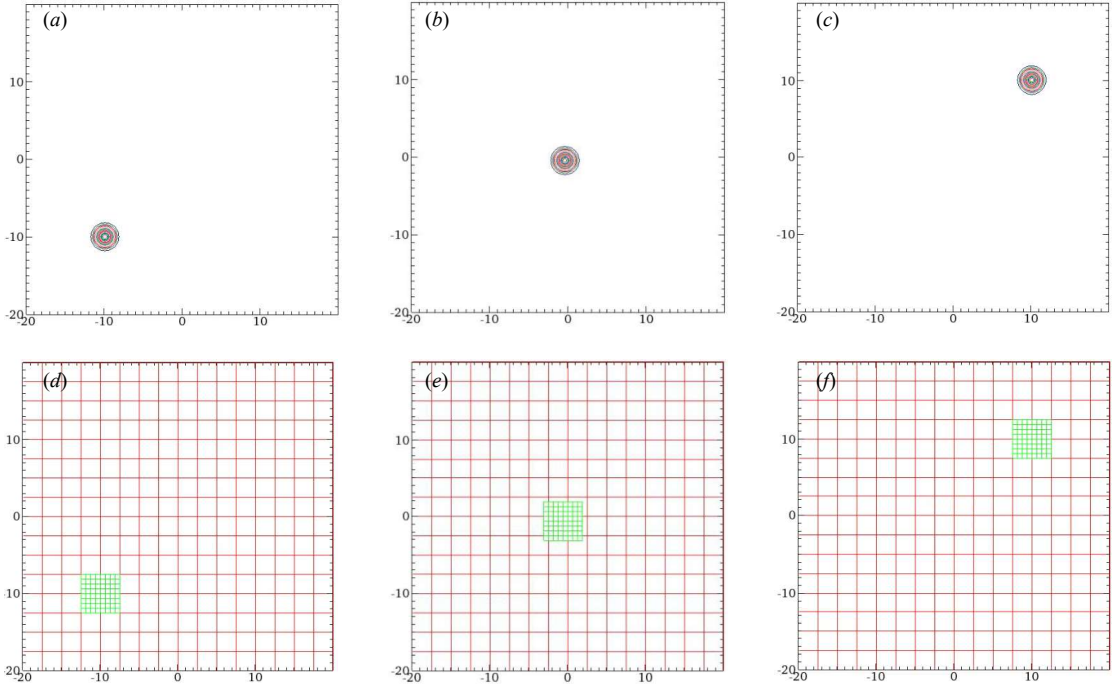

Figure 16: Numerical maps of (a-c) pressure contours and (d-f) multi-level grids in the two-dimensional SAMR isentropic vortex convection case.

*5.3.4. Two-dimensional Riemann Problem*

To further test this solver for handling multidimensional discontinuity, we consider the two-dimensional Riemann problem with the initial solution

$$(\rho, u, v, p) = \begin{cases} (1.138, 1.206, 1.206, 0.029), & (x, y) \in (0, 0.8] \times (0, 0.8] \\ (0.5323, 1.206, 0, 0.3), & (x, y) \in (0.8, 1) \times (0, 0.8] \\ (0.5323, 0.1, 1.206, 0.3), & (x, y) \in (0, 0.8] \times (0.8, 1) \\ (1.5, 0, 0, 1.5), & (x, y) \in (0.8, 1) \times (0.8, 1). \end{cases} \quad (43)$$

The simulation employs a two-dimensional domain of size $1.0 \times 1.0$. The computational domain is discretized into a grid of 448×448 cells, with a maximum of two additional refinement levels. Boundary conditions in both *x*- and *y*- directions are set to be outflow. The simulation employs a time-space third-order method and the HLLC Riemann solver. The Euler equations are solved without the viscous terms, using a minmod limiter during flux reconstruction. The simulation runs for a physical time of 1.0 s, at a CFL number of 0.99. The interested flow features are refined based on gradient criterion of density, pressure, and temperature.

It is observed in Fig. 17 that our spatiotemporally third-order method performs well in preserving the symmetry of the basic structures after the release of the initial discontinuity. The SAMR also performs well in recognizing subtle structures such as the unstable slip lines and inversed vortex, with these structures all well captured with the finest grid at a maximum level.

This canonical problem is well validated in both CPU and GPU, with their performance comprehensively compared in Table 3 earlier in Section 4.2. Overall, we have made considerable speedup on this example by using our parallelization strategy in Section 4.1.

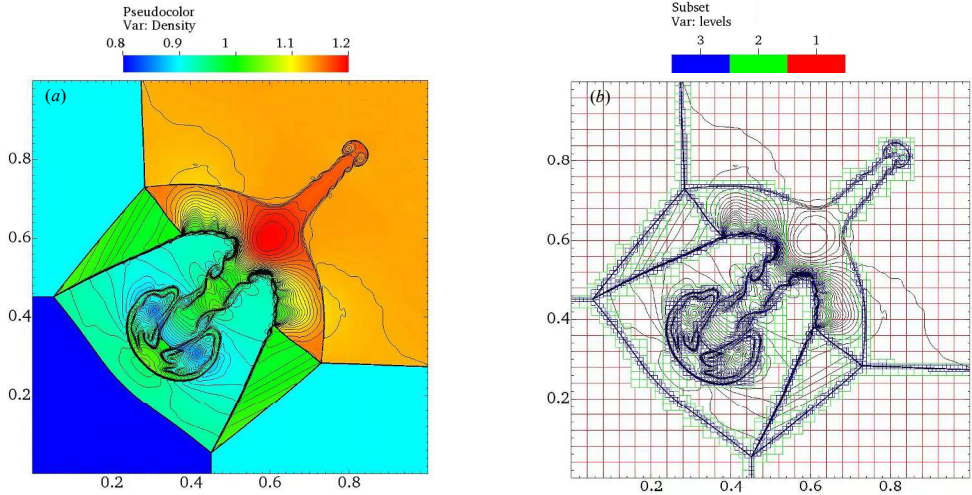

Figure 17: Numerical maps of density pseudocolor and multi-level grids superimposed with the density contour in the two-dimensional Riemann problem.

*5.3.5. Two-dimensional Cellular Detonation Propagation*

The simulation focuses on a two-dimensional detonation problem in a domain of 8 cm × 3 cm. The initial reactive mixture is a hydrogen-oxygen-argon system with a molar ratio of $H_2 : O_2 : Ar$ = 2:1:7. This system with a large portion of argon dilution has been validated in experiments to produce very regular detonation cells and thus was widely employed for numerical validation [6]. The initial pressure and temperature of the mixture are 6.67 kPa and 298 K, respectively. A one-dimensional ZND structure is superimposed, with the leading shock positioned at $x$ = 3 cm. Parameters of the one-dimensional ZND structure obtained from the Burke et al's model is listed in Table 13. This chemistry model is known to give better prediction of detonation cells among numerical simplified chemistry models, in alignment with the experiments. A pocket of unburned gas with a width of 1 cm and a height of 0.7 cm is introduced at $d$ = 0.3 cm behind the leading shock, at a pressure of 5.e5 Pa and a temperature of 2500 K, to perturb the ZND structure and initialize the cellular detonation.

The two-dimensional computational domain is discretized into 160×64× cells, with a maximum of two additional refinement levels. The simulation uses a time-space second-order accurate method and a CFL number of 0.8. The AUSM solver is employed with a van Albada limiter, with the chemistry enabled. The simulation runs until a physical time of 1e-4 seconds is reached. Boundary conditions are set as the inflow at the left and outflow on the right, with symmetry/adiabatic slip wall condition at the top and bottom boundaries. Furthermore, the smoke foil is obtained by tracking the triple point trajectory by pressure gradient.

Table 13: Parameters of the one-dimensional ZND structure obtained using Burke's 13-species 27-step chemistry model.

| | |
|---|---|
| $D_{CJ}$ [m/s] | 1616 |
| $p_{vN}$ [kPa] | 174 |
| $u_{vN}$ [m/s] | 393.89 |
| $T_{vN}$ [K] | 1900.1 |
| $t_{ig}$ [$\mu$s] | 3.875 |
| $\delta_{ig}$ [mm] | 1.532 |

Figure 18 and 19a illustrates the numerical results obtained on GPU, with the results on CPU quite similar. The results agree well with those obtained from previous literature. It is observed that both the gas-dynamic structures and the reaction front are well-captured and refined to a maximum level. The second refinement level grids take rough 3/4 portion of the entire domain, mostly because we set the minimum block size quite large to saturate the computing capacity of GPU. This strategy may cause additional refinement but could be quite efficient on GPU, since it significantly reduces the launching of device kernels by combine many small boxes into a big one. It is also illustrated in Fig. 19b that our solver generates regular detonation cells and the numerical smoke foil coincides well with the experimental results with the width of detonation cells roughly as $\Omega \approx 3$ cm. Table 14 shows the optimal computation performance on a 18-core Intel i9 CPU and one a NVIDIA V100 GPU. As can be seen, a considerable 6.49× speedup is achieved.

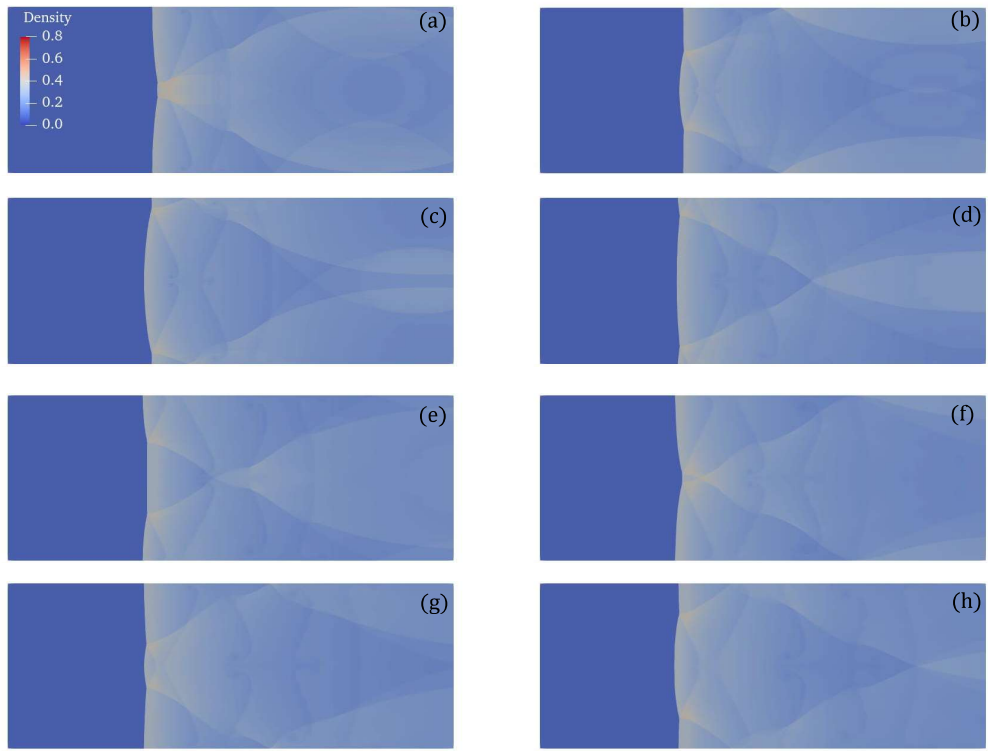

Figure 18: Pseudocolor density plots illustrating the evolution of dynamic cellular detonation structures.

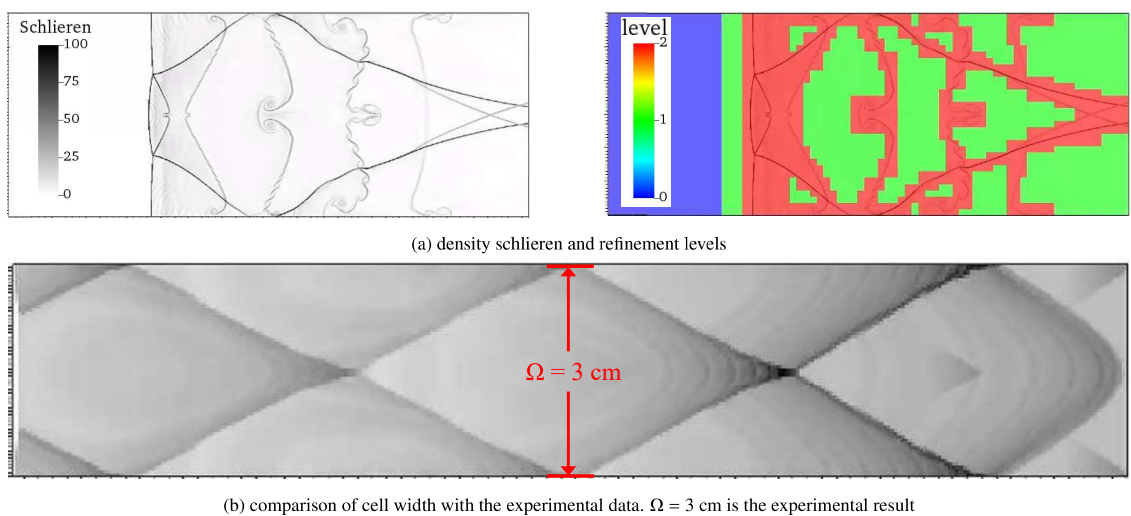

(a) density schlieren and refinement levels

(b) comparison of cell width with the experimental data. $\Omega = 3$ cm is the experimental result

Figure 19: Validation of adaptive mesh refinement on flow discontinuities and prediction on the detonation cell size with the conditions employed in the experiments [56].

Table 14: Performance of CPU and GPU on the 3-level 2-D cellular detonation problem.

|  | CFL number | Wall Time [s] | Speedup |
|---|---|---|---|
| i9-10980XE | 0.5 | 20340 | – |
| V100 | 0.5 | 4672 | 6.49 |

## 6. Application with multiple GPUs

This three-dimensional case is used to demonstrate solver's efficiency in performing large-scale DNS using the combination of SAMR and GPU-accelerated computation. The problem contains a rightward-traveling normal shock impacting and interacting with a mixture bubble compromising with stoichiometric H2 and O2 gases, as illustrated in Fig. 20. The reference results are from the uniform-grid DNS study by Diegelmmann et al [57], in which an optimal six-order WENO scheme [58] is applied. The initial and boundary conditions are discussed in details in [57].

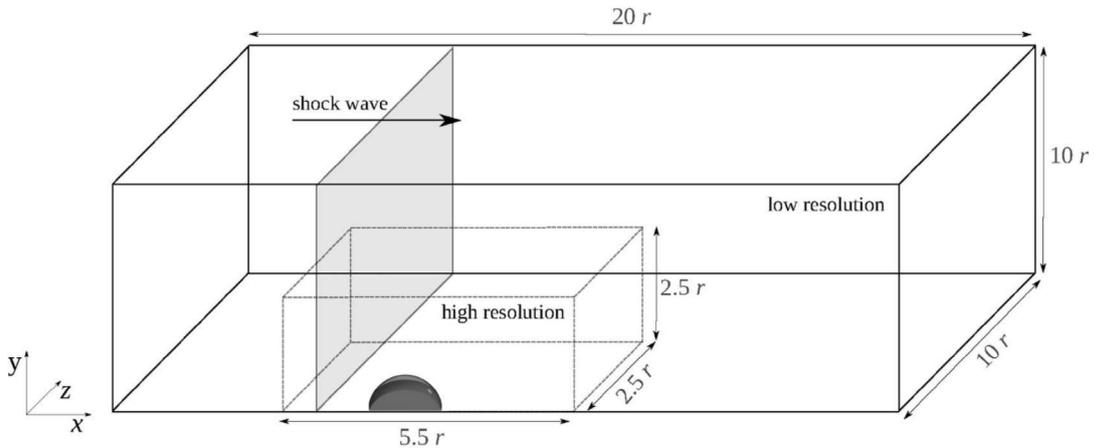

Figure 20: Schematic diagram of the RSBI problem from Diegelmann et al. [57].

Our SAMR study is performed on eight NVIDIA V100 GPUs, by employing the spatiotemporally second-order scheme introduced in Section 2.2. To acquire a similar grid resolution to that of Diegelmann et al.'s DNS study, we employ a total of four levels with refinement ratios $r_1 = 2, r_2 = 2, r_3 = 4$. By employing a base grid of dimensions $196 \times 96 \times 96$, we construct a finest grid size of 153.6 points per radius (pts/$r$), with $r$ the radius of the initial bubble. Other numerical conditions remain identical.

The computation is performed until $t$ = 5e-4 s, while we mainly focus here on the early development stages where the bubble is compressed and ignited to detonation by shock-bubble interaction, because during these stages there are most abundant results for us to conduct comparative analysis. The dynamic animation could be found in AMReX gallery (https://amrex-codes.github.io/amrex/gallery.html). Figure 21 depicts the evolution processes of the bubble and the reaction wave. It is observed that the stoichiometric H2-O2-Xe bubble is compressed and heated due to the shock impaction, leading to a direct ignition at about t = 42 $\mu$s, fitting well with the observation in Diegelmann et al.'s study. As the combustion wave propagates in the form of a burning ring through the bubble gas, by about t = 52 $\mu$s, the reaction wave has consumed the entire bubble gas, forming a bubble-shaped high-temperature region, as shown in Fig. 21(b-d). Subsequently, as seen in (e-h), the formation of the main vortex ring and the shedding of secondary vortices create the characteristic jellyfish-like structure typical to this three-dimensional reactive shock-bubble interaction (RSBI) problem.

Figure 22 presents two-dimensional slices in the $x - y$ plane at $z = 0$ during the interaction period. The upper images show the numerical schlieren overlaid with refined mesh configuration, while the lower images display the corresponding temperature superimposed with the bubble profiles. To discuss first, it can be seen that the SAMR algorithm performs very well, imposing the highest level of refinement to the leading shock wave, reflected shock, transverse waves and the bubble, while maintaining the base-level grid in the far-field regions. Furthermore, the results here demonstrate good agreement with those of the direct numerical simulation (DNS) study in Diegelmann et al. as well particularly in capturing the detonation waves in Fig. 22(a) and accurately reproducing the formation of the main vortex ring in (b) and the shedding process of secondary vortices in (c-d). By comparing the density flow field at $t = 70$ $\mu$s, our SAMR-DNS result agrees well with that from [57] qualitatively in details of the gasdynamic wave structures and bubble profile, as shown in Fig. 23a.

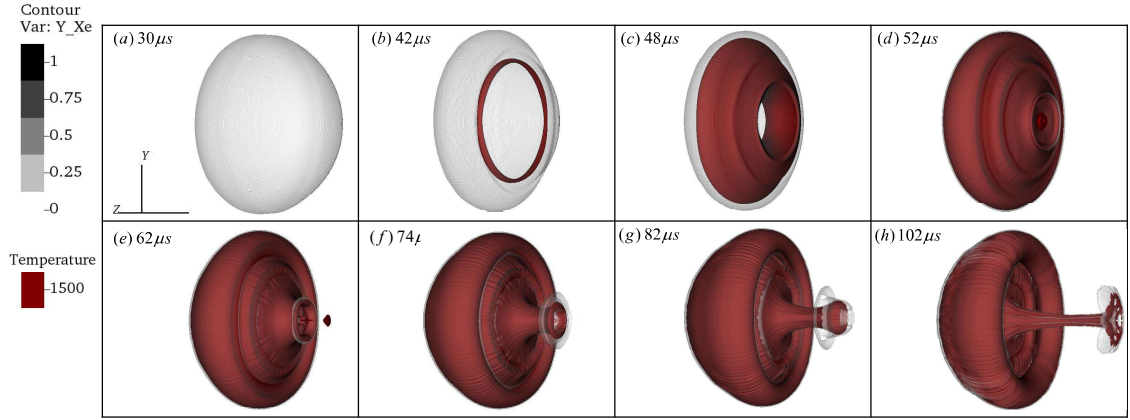

Figure 21: Three-dimensional RSBI contours of Xe mass fraction and temperature illustrating the reaction ring and jellyfish-like structures with an incident shock Mach number of $Ma = 2.83$.

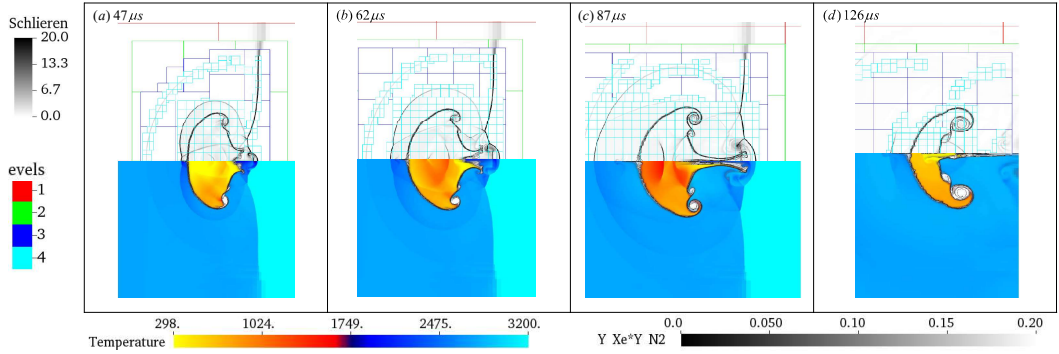

Figure 22: Two-dimensional RSBI slices of detailed fluid structures and adaptive mesh distributions with an incident shock Mach number of $Ma = 2.83$.

To investigate the induction and ignition processes, we also compare the temporal evolution of maximum radical species concentrations and temperature during RSBI with the DNS results of Diegelmann et al., as illustrated in Fig. 23b. The comparison shows basically consistency, although our SAMR-DNS results exhibit about 10 $\mu$s-period delay of the onset of detonation and about one order of magnitude lower mass fractions of the intermediate radicals. We analyzed this mostly attributed to the numerical dissipation that exists in the current second-order discretization method that restrains the capturing resolution of chemical process. And it would be significantly improved by implementing a higher order spatial scheme like the WENO schemes, which will be done in our further investigation. In addtion to that, it is noticeable that a total of maximum 53.6 million cells (about 50% reduction in Diegelmann et al.'s DNS [57]) is achieved via the use of adaptive mesh refinement, enabling significantly high computational efficiency within one day to complete this case. This result significantly highlights the efficiency of SAMR computation with GPU acceleration.

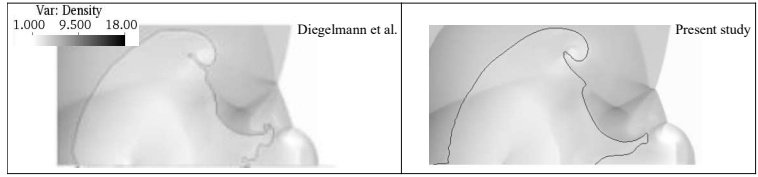

(a) Comparison of density flow field on a 2D slice at $z = 0$ and $t = 70~\mu s$

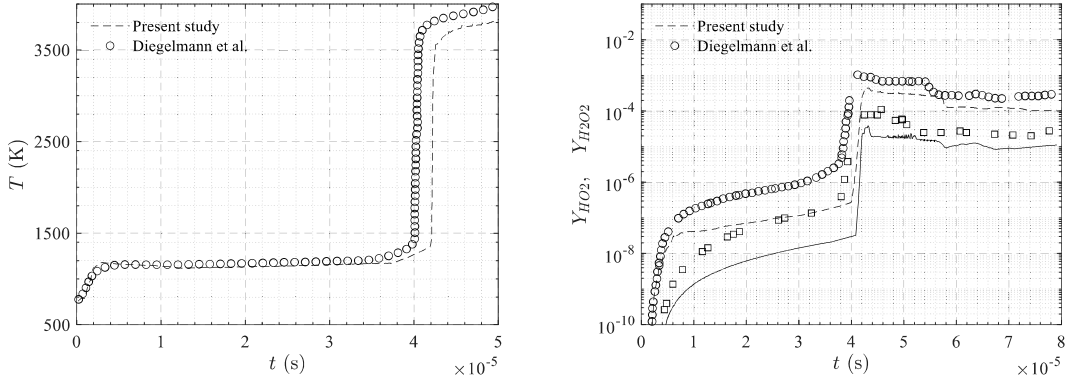

(b) Comparisons of temporal evolutions of the maximum temperature and intermediate species mass fractions of the entire domain

Figure 23: Validation of the present syudy against Diegelmann et al.'s DNS study [57]: (a) comparison of density flow field distributions at $t = 70~\mu s$, (b) comparisons of temporal evolutions of the maximum temperature and mass fractions of intermediate species ($Y_{HO_2}$ and $Y_{H_2O_2}$) during the shock impacting and detonation transition stages of RSBI.

## 7. Conclusion

This work presents a high-fidelity GPU-accelerated SAMR framework for compressible reactive flows, making several significant contributions to the field of computational fluid dynamics. Our implementation successfully addresses the fundamental challenges of implementing AMR algorithms on GPU architectures, demonstrating that the inherent tension between AMR's irregular data structures and GPUs' preference for regular computations can be effectively managed through careful algorithm design and optimization.

A key innovation lies in our elaborate time-stepping strategy, which supports both subcycling and non-subcycling approaches, providing flexibility in numerical integration while strictly maintaining conservation properties through a specialized refluxing algorithm. This algorithm, designed for arbitrary-order multi-stage Runge-Kutta temporal schemes, ensures solution accuracy across refinement boundaries, as validated through extensive numerical tests.

For chemical integration, we introduce a GPU-optimized low-storage explicit Runge-Kutta method that significantly improves performance through reduced register usage. This optimization proves particularly effective for both thread-coherent and thread-divergent scenarios, demonstrating the framework's ability to efficiently handle diverse computational patterns.

The framework's performance characteristics are thoroughly documented through comprehensive validation and scaling studies. Systematic comparisons between CPU and GPU implementations demonstrate remarkable speedups across various computational scenarios, with consistent advantages on both uniform and adaptive grids. The framework maintains high performance even when handling computationally demanding problems that involve complex physics and spatio-temporally distributed stiffness like the detonation propagation. Moreover, weak scaling tests confirm excellent parallel efficiency across multiple GPU nodes; The framework's capability for large-scale scientific computing applications is further demonstrated via a three-dimensional RSBI problem.

Future work will focus on to further enhance the framework's capabilities. First, we plan to incorporate higher-order numerical schemes into the SAMR framework, which requires careful treatment of the synchronization between fine and coarse grid solutions to maintain both accuracy and conservation properties. Second, we aim to incorporate neural networks and machine learning techniques in compressible reactive flow simulations for acceleration of the chemical source term evaluations, construction of reduced-order models, and improvement of closure relations for turbulence-chemistry interactions. These improvements will be crucial for simulating complex chemical systems with larger reaction mechanisms while maintaining computational efficiency.